\documentclass[%
 preprint, 
superscriptaddress,
 amsmath,amssymb,
 aps, physrev,
]{revtex4-2}

\usepackage{graphicx}
\usepackage{dcolumn}
\usepackage{bm}
\usepackage{hyperref}

\setcitestyle{super}

\usepackage{xr}

\usepackage[usenames,dvipsnames]{color}
\usepackage{colortbl}
\usepackage[english]{babel}
\newcommand{\DEL}[1]{}

\begin{document}


\title{Two-color harmonic spectroscopy of ultrafast Dirac electron dynamics}

\author{Zhaopin Chen}
\altaffiliation{These authors contributed equally to this work.}
\affiliation{Department of Physics, Technion -- Israel Institute of Technology, Haifa 3200003, Israel}
\affiliation{Solid State Institute, Technion -- Israel Institute of Technology, Haifa 3200003, Israel}
\affiliation{The Helen Diller Quantum Center, Technion -- Israel Institute of Technology, Haifa 3200003, Israel}

\author{Camilo Granados}
\altaffiliation{These authors contributed equally to this work.}
\affiliation{Department of Physics, Technion -- Israel Institute of Technology, Haifa 3200003, Israel}
\affiliation{Department of Physics, Guangdong Technion -- Israel Institute of Technology, Shantou 515063, Guangdong, China}
\affiliation{Guangdong Provincial Key Laboratory of Materials and Technologies for Energy Conversion, Guangdong Technion -- Israel Institute of Technology, Shantou 515063, Guangdong, China}
\affiliation{Eastern Institute of Technology, Ningbo 315200, China}

\author{Eyal Uzner}
\affiliation{Department of Physics, Technion -- Israel Institute of Technology, Haifa 3200003, Israel}
\affiliation{Schulich Faculty of Chemistry, Technion -- Israel Institute of Technology, Haifa 3200003, Israel}

\author{Ido Nisim}
\affiliation{Department of Physics, Technion -- Israel Institute of Technology, Haifa 3200003, Israel}
\affiliation{Solid State Institute, Technion -- Israel Institute of Technology, Haifa 3200003, Israel}
\affiliation{The Helen Diller Quantum Center, Technion -- Israel Institute of Technology, Haifa 3200003, Israel}

\author{Daniel Kroeger}
\affiliation{The Norman Seiden Multidisciplinary Graduate Program in Nanoscience and Nanotechnology, Technion -- Israel Institute of Technology, Haifa 3200003, Israel}
\affiliation{Solid State Institute, Technion -- Israel Institute of Technology, Haifa 3200003, Israel}
\affiliation{The Helen Diller Quantum Center, Technion -- Israel Institute of Technology, Haifa 3200003, Israel}

\author{Ofer Neufeld}
\affiliation{Schulich Faculty of Chemistry, Technion -- Israel Institute of Technology, Haifa 3200003, Israel}

\author{Marcelo F. Ciappina}
\affiliation{Department of Physics, Technion -- Israel Institute of Technology, Haifa 3200003, Israel}
\affiliation{Department of Physics, Guangdong Technion -- Israel Institute of Technology, Shantou 515063, Guangdong, China}
\affiliation{Guangdong Provincial Key Laboratory of Materials and Technologies for Energy Conversion, Guangdong Technion -- Israel Institute of Technology, Shantou 515063, Guangdong, China}

\author{Michael Krüger}
\affiliation{Department of Physics, Technion -- Israel Institute of Technology, Haifa 3200003, Israel}
\affiliation{Solid State Institute, Technion -- Israel Institute of Technology, Haifa 3200003, Israel}
\affiliation{The Helen Diller Quantum Center, Technion -- Israel Institute of Technology, Haifa 3200003, Israel}

\begin{abstract}
{\bf High-harmonic generation (HHG), the hallmark effect of attosecond science, is a nonperturbative nonlinear process leading to the emission of high-harmonic light from gases and solids. In gases, extreme driving laser pulse intensities can deplete the ground state, suppressing harmonic emission during the trailing edge of the pulse. Here, we report a similar effect, pronounced ultrafast carrier saturation dynamics and harmonic emission suppression during nonperturbative harmonic generation (NPHG) in a gapless Dirac semimetal—highly oriented pyrolytic graphite (HOPG). Remarkably, HOPG supports NPHG at laser intensities as low as $\sim 10^{10}$ W\,cm$^{-2}$, facilitated by its vanishing bandgap. Ultrafast carrier saturation strongly modulates the interplay between interband and intraband currents, a key characteristic of NPHG in Dirac materials. Using two-color spectroscopy, we reveal the excitation dynamics of Dirac electron-hole pairs as it affects the emission of harmonics during the presence of the driving laser pulse. The excitation of out-of-equilibrium hot carriers and the concomitant saturation near the Dirac points leads to a marked suppression of interband harmonics and induces measurable temporal shifts. These observations are supported by simulations based on semiconductor Bloch equations. Our finding reveal that field-driven carrier saturation plays a critical role in gapless solid NPHG. We demonstrate the potential of NPHG and HHG as a sensitive, all-optical probe of ultrafast carrier dynamics, offering novel opportunities for ultrafast optoelectronics in Dirac materials.}
\end{abstract}

\maketitle

\section*{Main}\label{sec1}  

Parametric nonlinear optical processes, such as second-harmonic generation or four-wave mixing, transiently modify the quantum state of a medium during the laser interaction, but leave it unchanged afterward~\cite{Boyd2008}. This framework extends even to high-harmonic generation (HHG), an extreme nonlinear field-driven process observed mainly in gases~\cite{Ferray1988,McPherson1987,Lewenstein1994} and condensed-matter systems~\cite{Ghimire2011,Vampa2015a,Ghimire2019,Borsch2023,Heide2024,chen2025attosecond}, which cannot be described by perturbative nonlinear optics. Although HHG itself is parametric, competition with other strong-field processes can affect HHG, most prominently photoionization~\cite{Lewenstein1994}. In gases, sufficiently intense and long laser pulses can substantially deplete the ground-state electron population, suppressing harmonic emission, causing spectral shifts and reducing harmonic cutoff energies~\cite{Schafer1997,Shin1999}. Additionally, the resulting high densities of photoelectrons adversely affect harmonic phase matching~\cite{Rae1994,Gaarde2008}.

In solids, analogous 
phenomena causing the suppression of the emission of harmonics might be expected at considerably lower laser intensities due to their smaller bandgaps relative to atomic ionization energies. However, experimental observation of pronounced 
carrier excitation dynamics during solid-state HHG and, more generally, nonperturbative harmonic generation (NPHG) remains elusive. Achieving a significant suppression of harmonic emission typically requires excitation fractions on the order of $\sim40\%$, while in experiment, typical excitation levels in semiconductors are usually limited to a few percent to avoid material damage~\cite{Suppression_nc2017}. Pre-excitation of carriers with ultraviolet pulses has been explored as an alternative route to reach a carrier saturation regime and suppress the harmonic 
yield in semiconductors~\cite{Suppression_nc2017,cheng2020ultrafast,heide2022probing}. However, such approaches obscure the intrinsic saturation dynamics driven solely by the fundamental laser pulse which drives the harmonic generation process. Here, the subject of saturation and harmonic yield suppression is an intricate one because condensed matter is far more complex than the gas phase due to many-body scattering and decoherence, lower symmetry, and Bloch state delocalization. Moreover, the microscopic origins of harmonic suppression in solids remain actively debated. Such suppression could arise predominantly from state blocking, where transitions from valence to conduction bands become unavailable due to occupied states, or from excitation-induced dephasing, where increased carrier densities accelerate electron scattering and degrade interband coherence~\cite{van2024toward}. 

Highly oriented pyrolytic graphite (HOPG), a structurally robust carbon-based bulk material composed of stacked graphene layers~\cite{Pappis1961,Blackman1962}, provides a promising platform for investigating such intrinsic saturation dynamics under laser-field driving. HOPG retains the key electronic properties of monolayer graphene, including its gapless Dirac dispersion, high carrier mobility, and strong anisotropic optical responses~\cite{CarSynth,HOPG,zhou2006first}. It also offers additional experimental advantages, such as structural robustness and scalability, that overcome the intrinsic limitation of graphene’s low damage threshold. Furthermore, HOPG allows systematic exploration of the effects of interlayer coupling and thickness variations on nonlinear optical and electronic dynamics~\cite{3rdHGraphe}. Despite these compelling features, HHG and NPHG of low-order harmonics in HOPG remain experimentally unexplored to date.

In this combined experimental and theoretical study, we introduce an all-optical two-color spectroscopy technique to study ultrafast population dynamics in the Dirac cone, revealing behavior beyond that observed in conventional gapped solids. Specifically, we observe a time shift in harmonic emission relative to the driving field peak, a clear signature of laser-field-induced carrier saturation in a gapless Dirac semimetal. Whereas earlier studies of graphite and graphene have probed photon-induced saturation, Pauli blocking, and relaxation dynamics using transient absorption spectroscopy~\cite{breusing2009ultrafast, brida2013ultrafast}, our work explores a fundamentally distinct regime, the nonperturbative regime, where carrier saturation evolves within tens of femtoseconds and directly influences harmonic generation. This bridges the gap between weak-field saturation effects and field-driven dynamics in Dirac materials, uncovering a new class of ultrafast phenomena governed by light-driven population reshaping. More broadly, the ability to resolve femtosecond-scale carrier saturation through HHG suppression suggests that this approach could also serve as a sensitive, all-optical probe of light-induced phase transitions in solids as demonstrated in both experimental~\cite{cavalleri2001femtosecond,bionta2021,nie2023following} and theoretical~\cite{wegkamp2015ultrafast,murakami2018high,silva2018high, valmispild2024sub} studies.

\section*{Results}

\subsection*{Nonperturbative harmonic generation in HOPG}\label{sec2}

\begin{figure}[!htb]
\centering
\includegraphics[width=1\textwidth]{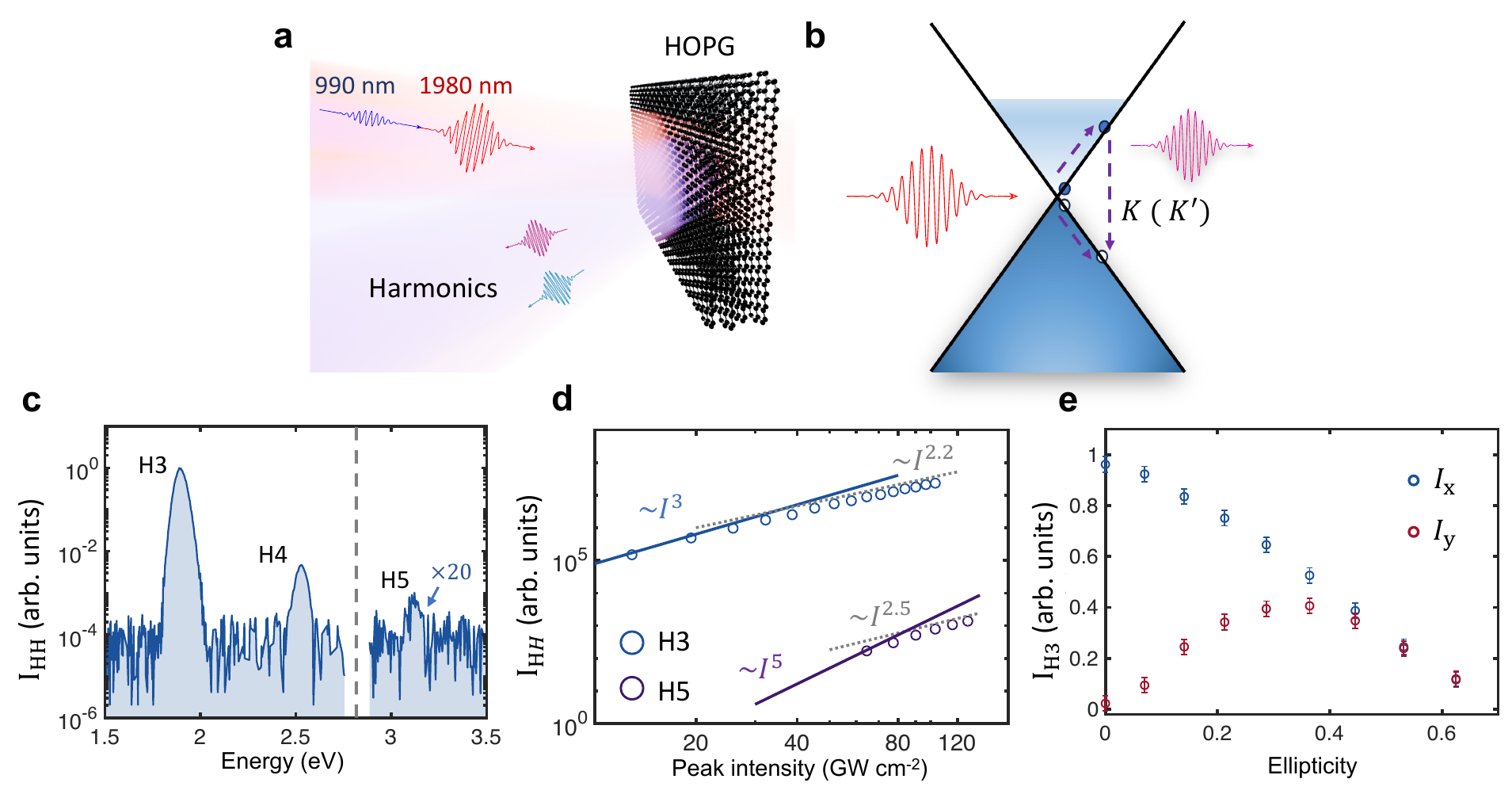}
\caption{\textbf{Nonperturbative harmonic generation from HOPG by a two-color pulse.}
{\bf a}, Conceptual sketch of nonperturbative harmonic generation in HOPG driven by a strong driving laser pulse with 1980 nm central wavelength and a weak second harmonic pulse. {\bf b}, Three-step model of interband HHG around the Dirac cone. {\bf c}, Experimentally measured harmonic spectrum with a laser peak intensity of 72 GW\,cm$^{-2}$ and $3\%$ second harmonic admixture. {\bf d}, Intensity scaling of the third and fifth harmonic yield when driven by the fundamental pulses only. The third harmonic exhibits a clear transition from a perturbative regime ($\propto I^{3}$) to a nonperturbative regime ($\propto I^{2.2}$), while the fifth harmonic is purely nonperturbative over the measured range, scaling as $I^{2.5}$ at the highest intensity. {\bf e}, Dependence of the harmonic intensity on laser ellipticity ($\epsilon = E_x/E_y$), for both $x$- and $y$-polarized components.} \label{fig1}
\end{figure}

We generate nonperturbative harmonics from an HOPG sample in reflection geometry~\cite{Vampa2018} using an infrared femtosecond pulse (frequency $\omega_0$) combined with a weak, synchronized second harmonic (SH) field ($2\omega_0$), with a polarization parallel to the fundamental pulse (see Fig.~\ref{fig1}a). The fundamental laser pulse has a central wavelength of 1980 nm and a duration of 60 fs (about 9 optical cycles). The SH field is generated by focusing the fundamental field onto a $\beta$-barium borate (BBO) crystal. Both beams are split and then recombined in an interferometric setup, achieving both spatial and temporal overlap with accurate control of the two-color ($\omega_0$--$2\omega_0$) time delay $\tau$ (see Methods). Throughout this work, we set the SH pulse intensity at 3\% of the fundamental laser intensity in both experimental measurements and theory calculations.
 
HOPG exhibits an AB-stacked multilayer graphene structure, resulting in a stable crystalline form whose band dispersion near the Dirac point closely resembles that of monolayer graphene. For comparison, ab initio calculations of the electronic band structures of both graphene and HOPG are provided in Fig.~S12 of the Supplementary Information. Since our HOPG samples are  undoped, the Fermi level lies close to the Dirac point ($\mu \approx 0$), consistent with ARPES measurements on graphite~\cite{zhou2006first}. At room temperature ($k_\mathrm{B}T \approx 26$ meV), any residual Fermi-level shift is small compared to the driving laser photon energy (0.63 eV). We therefore assume $\mu \approx 0$ for both the experiment and the theoretical modeling. Given the moderate laser field strength used in our experiments -- limited by the damage threshold of the material -- the electron dynamics are effectively confined to the vicinity of the Dirac cone (Fig.~\ref{fig1}b). In the nonperturbative regime relevant to our work, the emission of above-bandgap harmonics is primarily governed by interband polarization processes \cite{Vampa2015a}. This behavior is further corroborated by numerical simulations based on the semiconductor Bloch equations (SBEs), which reveal a clear predominance of interband contributions in the generated harmonic spectrum (see Fig.~S5 in Supplementary Information). Interband polarization roughly corresponds to the three-step picture of HHG. Here, interband tunneling produces an electron-hole pair that is driven by the laser field in momentum space until electron and hole recombine and a harmonic photon is emitted~\cite{Vampa2015a}.

In our experiment, nonperturbative harmonics are generated from HOPG with spectral features observed up to the fifth order (3.1\,eV photon energy), exceeding the previously reported harmonic cutoff in monolayer graphene of about 2.4\,eV~\cite{ExoticGraph2,taucer2017nonperturbative}. Figure~\ref{fig1}c displays the measured harmonic spectrum obtained using the 1980\,nm driving pulse with a peak intensity of 72\,GW\,cm$^{-2}$ and the admixed SH pulse. With respect to the third harmonic, the fourth harmonic exhibits a relative intensity of $I_{\mathrm{H}4}/I_{\mathrm{H}3} = 6.3\%$, while the second harmonic is omitted from the spectrum, as it is masked by the much more intense incident SH field.

To confirm the nonperturbative nature of the harmonic emission, we measure the intensity scaling of the third harmonic (H3) driven by a single-color 1980\,nm pulse (Fig.~\ref{fig1}d). At low laser intensities, H3 follows the expected perturbative scaling of $I^3$, but with increasing intensity, the scaling gradually transitions to $I^{2.2}$. In the same intensity range, the fifth harmonic (H5) emerges and exhibits a distinctly nonperturbative scaling of approximately $I^{2.5}$ at the highest intensity, rather than the perturbative $I^5$ dependence. These clear deviations from perturbative scaling marks the onset of NPHG~\cite{taucer2017nonperturbative,ExoticGraph2,cha2022gate}. In this regime, the nonlinear response arises primarily from carrier acceleration and recombination rather than from conventional multiphoton transitions~\cite{ExoticGraph2,Avetissian2022}. Moreover, the harmonic yield shows a nontrivial dependence on the laser ellipticity (Fig.~\ref{fig1}e), consistent with a defining signature of graphene NPHG observed at longer wavelengths in experiment and theory~\cite{ExoticGraph2,liu2018driving,mrudul2021high,cha2022gate}. In particular, the $y$-polarized component is strongly enhanced under elliptically polarized excitation, reaching a maximum around $\epsilon \approx 0.35$, reflecting strong coupling between interband and intraband currents. 
Additionally, we observe a blue shift of the harmonic spectrum under strong excitation \cite{baudisch2018ultrafast}, reminiscent of the spectral shifts reported in atomic HHG \cite{gaarde1998macroscopic} (see Fig.~S13 in the Supplementary Information).

\subsection*{Emission time shift of harmonic generation}\label{sec3}

\begin{figure}[!htb]
\centering
\includegraphics[width=0.9\textwidth]{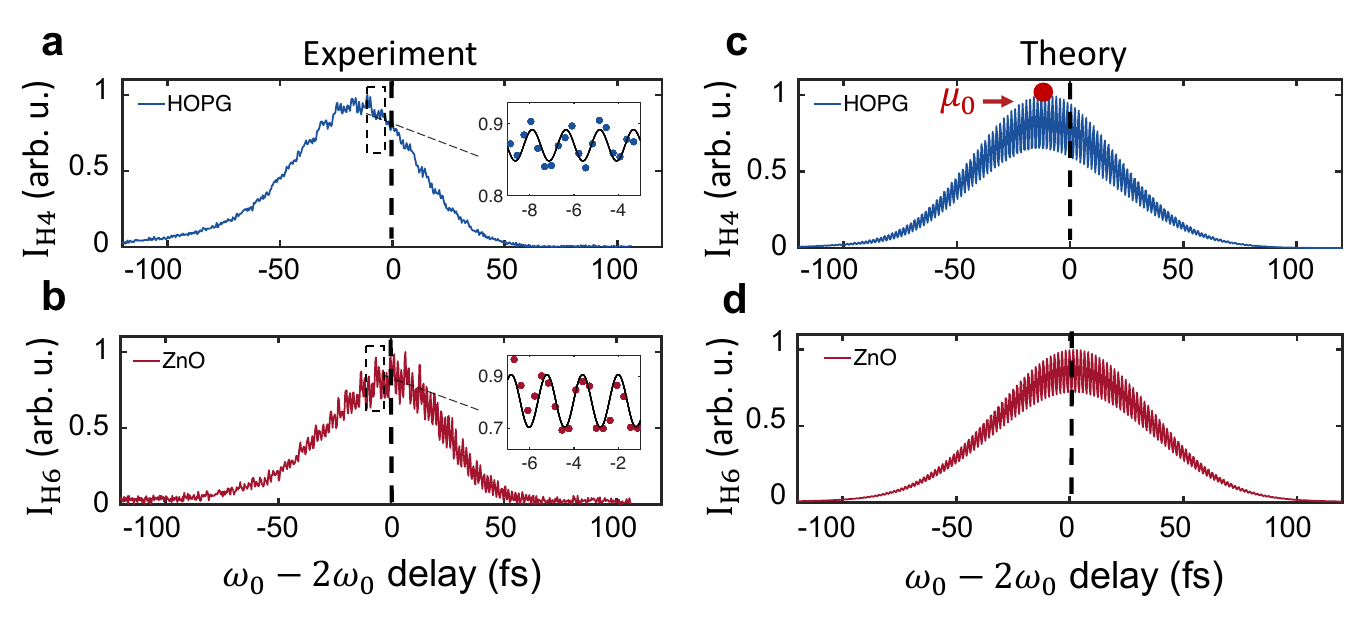}
  \caption{\textbf{Harmonic intensity as a function of $\omega_0$--$2\omega_0$ delay.}
  \textbf{a}, Experimentally measured harmonic intensities as function of the $\omega_0$--$2\omega_0$ delay for the fourth harmonic (H4) in HOPG with $\omega_0$ peak intensity $72\, \mathrm{GW cm^{-2}}$. \textbf{b}, The same for the sixth harmonic (H6) in ZnO with a laser intensity of $430\,\mathrm{GW cm^{-2}}$.
  \textbf{c} and \textbf{d}, Corresponding theoretical results simulated based on the semiconductor Bloch equations (Eq.~\ref{SBE3_P}) for comparison with the experimental data of HOPG and ZnO, respectively.
  Note that for negative delays, the $2\omega_0$ pulse arrives earlier at the harmonic generation medium than the $\omega_0$ pulse. The peak emission of H4 occurs approximately $17.5 \pm 0.2$~fs before the optimal temporal overlap in the experiment and $\sim$ 17~fs in the simulation.}
  \label{fig2}
\end{figure}

Two-color spectroscopy is a powerful all-optical approach capable of retrieving the temporal dynamics of the harmonic emission~\cite{NDudovich,uzan2022observation}. Here, we measure the harmonic intensity as a function of the relative time delay $\tau$ between the two colors ($\omega_0$–$2\omega_0$), as shown in Fig.~\ref{fig2}. The fourth harmonic (H4) is generated in HOPG when the $\omega_0$ and $2\omega_0$ pulses temporally overlap, enabled by symmetry breaking of the fundamental $\omega_0$ electric field through perturbation by the $2\omega_0$ pulse (Fig.~\ref{fig2}a). As a reference, we replace the HOPG sample with a conventional bulk gapped solid HHG sample, a ZnO crystal~\cite{Ghimire2011,Vampa2015a,tritschler2003evidence}, and measure the above-band gap sixth harmonic (H6) intensity as a function of $\tau$. The measurements in ZnO are performed under the same conditions as in HOPG, except that the fundamental laser intensity is increased to 384~GW cm$^{-2}$ to ensure the generation of above-band gap harmonics through the interband process. The band gap of the ZnO crystal is around 3.3 eV.  Full delay-dependent harmonic spectra are shown in Fig.~S2 of the Supplementary Information.

In each of the plots shown in Fig.~\ref{fig2}, oscillations of the harmonic signal with a frequency of 4$\omega_0$ are visible on top of an overall Gaussian-like curve. In our work, we focus on the latter. Remarkably, we observe that the maximum signal of H4 in HOPG occurs approximately $17.5 \pm 0.2$ fs before the temporal overlap of the two pulse intensity peaks at $\tau = 0$. In contrast, the H6 signal in ZnO exhibits a symmetric delay dependence, with its maximum occurring precisely at zero delay.  A symmetric behavior is also found for monolayer WS$_2$, a hexagonal gapped material (see Supplementary Information). For HOPG, the time shift $\mu_0$ corresponding to the maximum harmonic signal is extracted by fitting the entire delay-dependent harmonic trace with a Gaussian function and identifying the peak position of the fit. 

To gain further insight into the underlying physics, we perform theoretical simulations based on both one-dimensional (1D) and two-dimensional (2D) semiconductor Bloch equations (SBEs, see Methods in the main text and Supplementary Information for details) using the same parameters as in the experiment for both HOPG and ZnO (Fig.~\ref{fig2}c and d; see also full spectra in Fig.~S6 of the Supplementary Information). We solve the SBEs to obtain the full time-dependent microscopic populations and interband coherences induced by the laser pulse. In contrast to semiclassical Boltzmann equation approaches~\cite{brida2013ultrafast}, we incorporate electron–electron and electron–phonon scattering via phenomenological relaxation and decoherence times $(T_1, T_2)$, so that population redistribution and decoherence are treated at the microscopic level. In this framework, an out-of-equilibrium “hot” carrier distribution naturally emerges during the laser–matter interaction without explicitly describing thermalization or introducing an effective electron temperature (cf.~\cite{baudisch2018ultrafast}).

Under a fixed interband decoherence time of $T_2 = 6.6$ fs, both the 1D and 2D models exhibit a negative delay for the harmonic maximum (note that the negative delay in the 2D simulation is sensitive to the choice of $T_2$, see Fig.~S11 in the Supplementary Information for details). Since the 1D SBE model reproduces the main features observed in the 2D simulations, including the negative time shift of the harmonic maximum, we adopt the simpler 1D model for most of the remainder of our analysis. The simulation results show good agreement with the experimental data. We accurately reproduce the time shift of the H4 signal in HOPG, yielding a similar shift $\mu_0 = -17$ fs with respect to $\tau = 0$, and correctly predict the peak of the H6 signal in ZnO at zero delay. Furthermore, we simulate the delay dependence of H4, H6 and H8 from ZnO, and in all cases the harmonic maxima occur near $\tau = 0$, confirming that the choice of H6 as reference does not affect the evaluation of the delay. In addition, supplementary calculations that coherently average the harmonic emission over the optical penetration depth in the $z$-direction (Figs.~S15–S16) show that depth-dependent interference does not create a negative delay in a gapped system and has only a minor effect in the gapless case, indicating that the observed temporal shift cannot be attributed to propagation or geometric averaging effects.

\begin{figure}[!htb]
\centering
\includegraphics[width=1.0\textwidth]{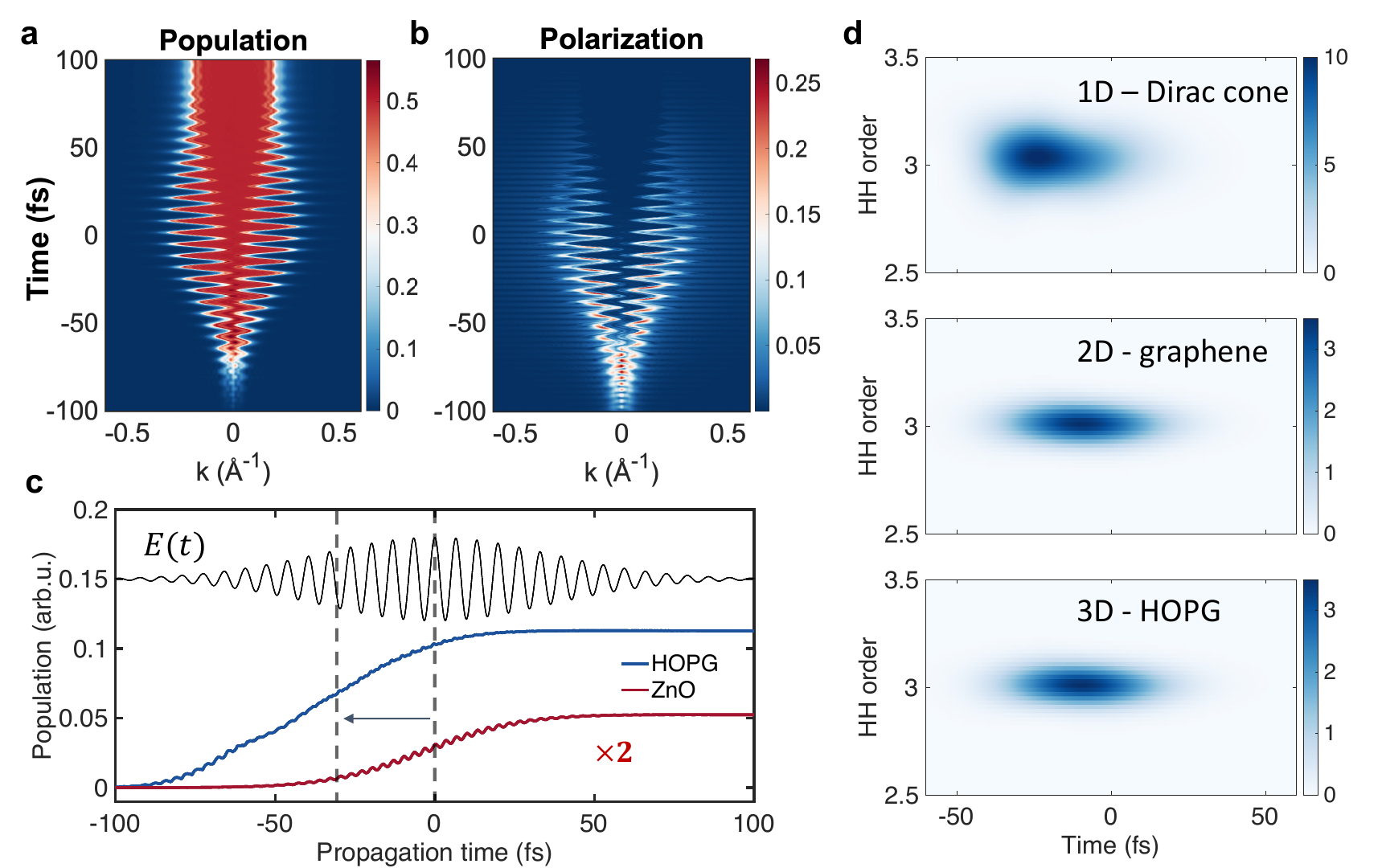}
 \caption{\textbf{Population saturation dynamics around Dirac points.}
  \textbf{a}, Calculated conduction band electron population dynamics and 
  \textbf{b}, interband polarization dynamics in momentum space during the excitation driven solely by the fundamental pulse at an intensity of $72\,\mathrm{GW\,cm^{-2}}$. In the gapless Dirac cone, carrier populations can reach up to 50\% at each $k$-point, leading to pronounced saturation effects.
  \textbf{c}, The total conduction band population dynamics show that the excitation of 50\% of the final population occurs well before the peak of the driving field. For comparison, the HHG process in ZnO at a peak intensity of $430\,\mathrm{GW\,cm^{-2}}$ shows that the excitation of 50\% of the final population coincides with the peak of the laser pulse.
  \textbf{d}, Time–frequency analysis of the interband harmonics, obtained via a Gabor transform for 1D (Dirac cone), 2D (graphene), and 3D (HOPG) models, reveals that the maximum intensity of the third harmonic (H3) occurs prior to the peak of the driving field.
}\label{fig3}
\end{figure}

To gain deeper insight into the pronounced time shift observed in NPHG from HOPG, we analyze the time evolution of both the conduction band electron population $f_c(k, t)$ and the interband polarization $p(k, t)$, as shown in Figs.~\ref{fig3}a and b. Our calculations reveal that this time shift originates from the strong interplay between intraband and interband currents -- a characteristic of NPHG in gapless Dirac materials that also underlies their nontrivial ellipticity response~\cite{al2014high,sato2021high,cha2022gate}. A significant accumulation of intraband population leads to a suppression of the interband polarization. Notably, the conduction band population near the Dirac point rapidly approaches half-filling well before the peak of the driving field, indicating a pronounced saturation effect. Unlike conventional solid-state harmonic processes where carrier saturation is negligible~\cite{GVampa2}, the time-dependent $k$-space populations $f_k^{c,v}(t)$ of conduction ($c$) and valence ($v$) bands play a crucial role in the interband transition term $\left( 1 - f_k^{c}(t) - f_k^{v}(t) \right) d^{c \nu}_k E(t)$ in the SBEs (see the Methods section, Eq.~(\ref{SBE3_P})). Once both $f_k^c$ and $f_k^v$ reach 0.5, the interband transition term vanishes, effectively blocking further interband excitation -- a phenomenon known as state blocking or state-filling saturation. This behavior is analogous to a two-level system where population inversion is forbidden and the excitation process saturates due to balanced electron occupation across bands.

The inability of the conduction band population to significantly exceed half-filling for most $k$-points around the Dirac point indicates that saturation occurs early during the fundamental driving pulse. This saturation not only limits further excitation, but also suppresses the interband polarization, which is an essential ingredient for interband harmonic generation. As a result, the dominant interband term is weakened, even near the peak of the driving pulse. Instead, the maximum interband harmonic intensity occurs prior to the onset of saturation.

To visualize this effect more clearly, we compute the \textit{total} conduction band population (around the cone) by integrating over momentum space, $n_c(t) = \int f_c(k, t)dk$, shown in Fig.~\ref{fig3}c. It is evident that the excitation of 50\% of the final total population (which remains after the end of the laser pulse) is achieved at a significantly earlier time than the peak of the driving field, underscoring the presence of strong saturation and depletion. In the absence of saturation, one would expect this time to coincide with the peak of the driving field due to its temporal symmetry. For comparison, we also plot the calculated total conduction band population dynamics in ZnO, which shows that the excitation of 50\% of the final total population occurs right at the peak field, consistent with the larger band gap that prevents efficient carrier saturation.

Furthermore, a time-frequency analysis using a Gabor transform (Fig.~\ref{fig3}d) shows that the maximum intensity of the H3 in HOPG driven by a single-color 2 $\mu$m pulse occurs before the field maximum ($t = 0$). In the 1D model this advance is approximately 25~fs, while in the 2D (graphene) model it is on the order of 10~fs, in agreement with a 3D (HOPG) model (see Supplementary Information). We note that there is a disagreement between these simulations on the exact prediction for the shift, none of which perfectly match with the experiment. However, the fact that all of these levels of theory get the correct trend is strong evidence that they indeed capture the essential physics. A complementary Gabor transform of the harmonic emission driven by $\omega_0$–$2\omega_0$ two-color pulses at various time delays (see Fig.~S8 in the Supplementary Information) demonstrates that the $2\omega_0$ field primarily acts as a probe, sampling the harmonic emission originating from the single-color $\omega_0$ excitation. This is in contrast to wide-band gap semiconductors where the non-depletion approximation is valid and interband polarization is largely unaffected by intraband population. In our case, the absence of a band gap in HOPG leads to substantial carrier excitation around the Dirac cone and strongly modifies the interband contribution to NPHG. Consequently, intraband population dynamics play a critical role in shaping the overall NPHG process in gapless Dirac materials such as HOPG. One can also expect to see a strong saturation effect in wide-bandgap semiconductor NPHG and HHG as long as the laser intensity is sufficiently large. However, the low damage threshold of these materials prevents a clear observation of this effect in gapped solids. Nonetheless, our simulations confirm that the harmonic emission time shifts significantly earlier than even in ZnO when its band gap is artificially reduced in simulations to a nearly gapless regime, as shown in Fig.~S9 of the Supplementary Information.

\begin{figure}[!htb]
\centering
\includegraphics[width=0.8\textwidth]{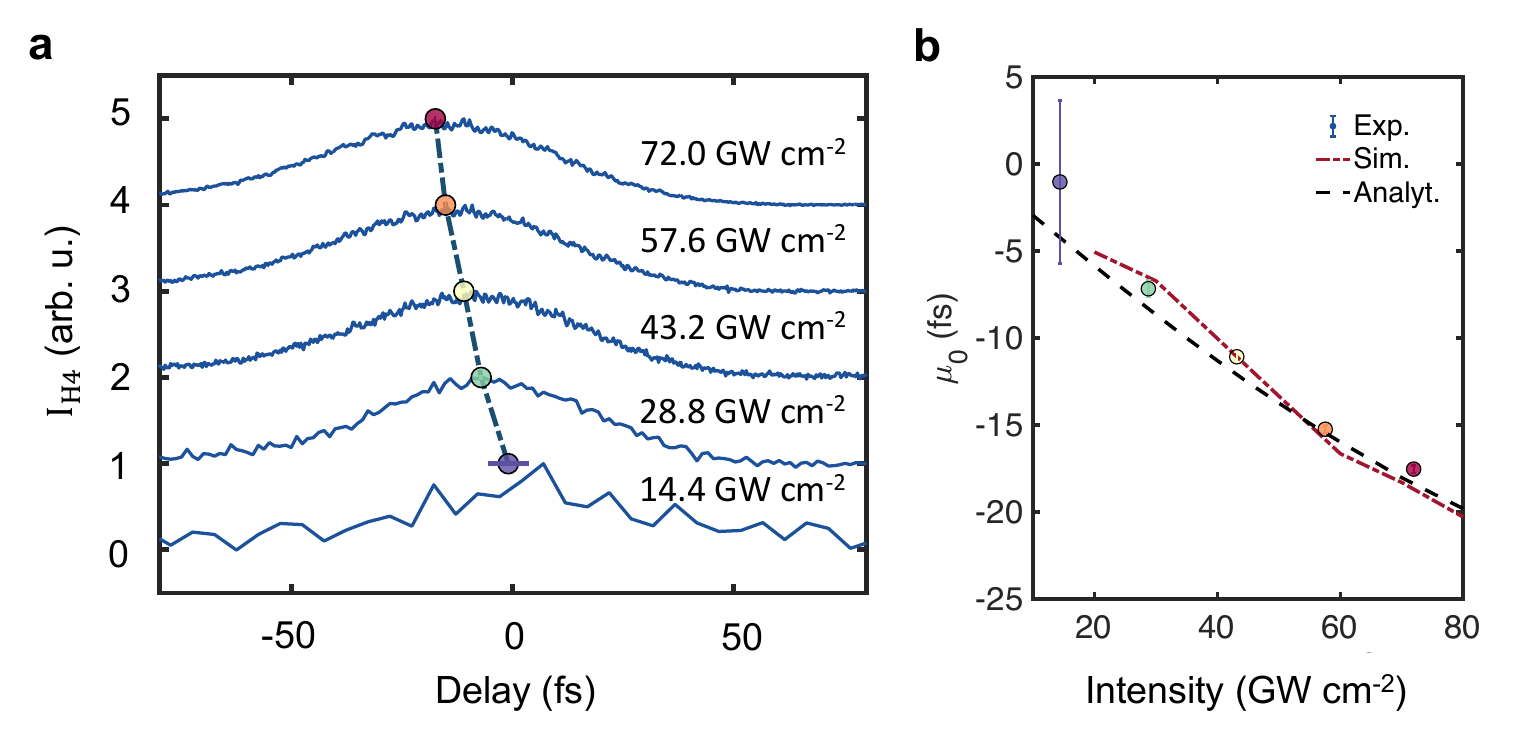}
 \caption{ \textbf{H4 emission time shift as a function of laser intensity.}  
\textbf{a}, Experimentally measured normalized H4 intensity as a function of $\omega_0$--$2\omega_0$ delay at different laser intensities. The relative delay position ($\mu_0$) of the maximum H4 intensity shifts to earlier times as the laser intensity increases. For better visibility, we introduce horizontal offsets.
\textbf{b}, Comparison of the maximum H4 intensity delay as a function of $\omega_0$ laser intensity obtained from experimental measurements, SBE calculations, and analytical results based on Eq.~\ref{analy_wt}. Experimental uncertainties for $\mu_0$ are smaller than the marker size at higher intensities.}  \label{fig4}
\end{figure}

Since the emission time shift originates from dynamic population saturation, a stronger driving laser intensity is expected to induce earlier saturation during the driving process. To investigate this, we experimentally measure the time shift of H4 under various $\omega_{0}$ laser intensities, as shown in Fig.~\ref{fig4}a. As the driving laser intensity increases, the peak of the H4 signal in the $\omega_0$--$2\omega_0$ delay scan systematically shifts to earlier times, further corroborating the presence of a saturation effect. The relationship between the H4 emission time shift $\mu_0$ and the driving laser intensity is plotted in Fig.~\ref{fig4}b.

Our SBE-based simulations accurately reproduce the experimental trend, confirming that the negative time shift $\mu_0$ increases with increasing peak intensity. To gain further insight into the phenomenon, we introduce an analytical approximation (Eqs.~\ref{Popul1}–\ref{analy_wt}), assuming non-depleted excitation from the valence band to the conduction band while accounting for the influence of electron occupation in both bands on the excitation rate. This model also predicts a similar dependence of $\mu_0$ on the laser intensity, as indicated by the black dashed line in Fig.~\ref{fig4}b.

\subsection*{Dirac electron population dynamics }

\begin{figure}[htb]
\centering
\includegraphics[width=1\textwidth]{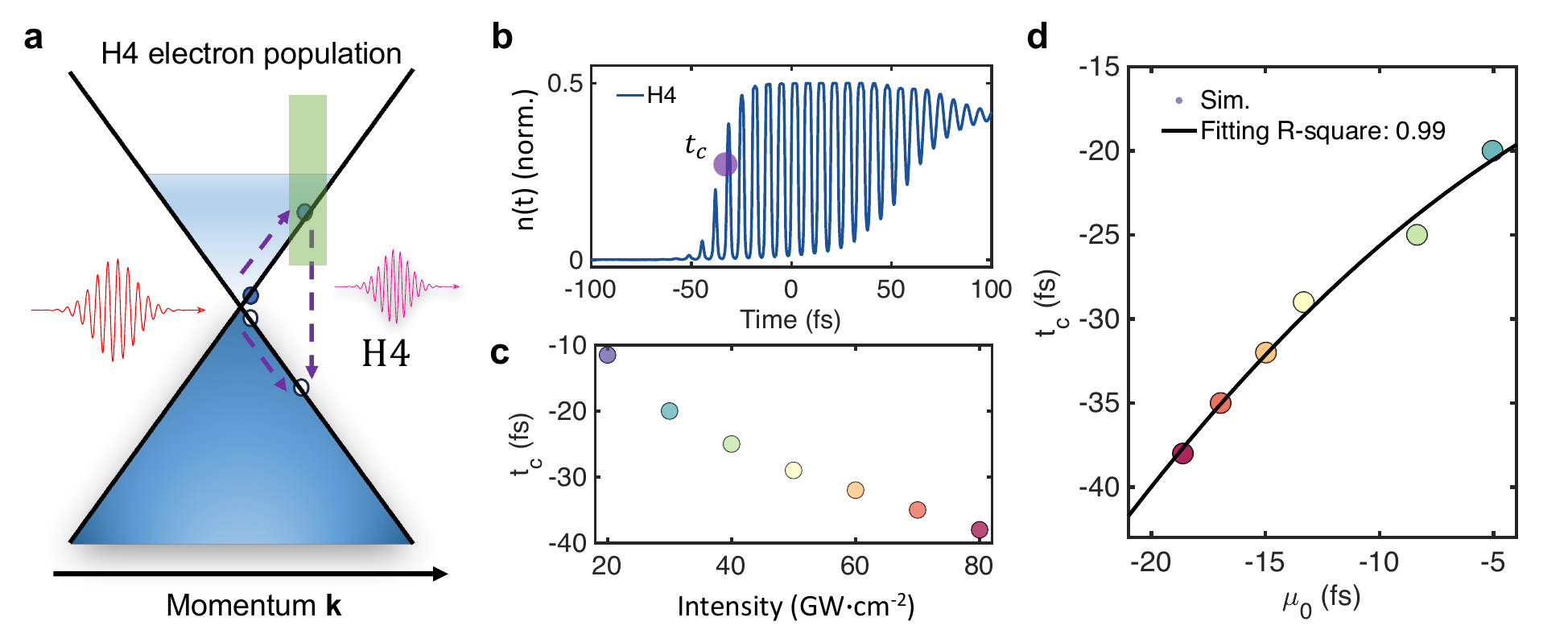}
 \caption{\textbf{Analysis of Dirac electron population dynamics via emission time shift.}  
\textbf{a}, Electron dynamics associated with fourth harmonic generation. The green shaded area indicates the region in momentum space where electron population contributes to H4 generation as the bandgap energy there corresponds to the H4 photon energy. 
\textbf{b}, Under excitation by a 2~$\mu$m pulse with a peak intensity of 70\,GW\,cm$^{-2}$, the H4 electron population oscillates as a function of propagation time. The time $t_c$ marks the moment when  a quarter of the electron population in that region are excited.  
\textbf{c}, The absolute value of $t_c$ increases with driving laser intensity, indicating an intensity-dependent excitation process.  
\textbf{d}, A strong correlation is observed between $t_c$ (in the single-color 2~$\mu$m excitation case) and $\mu_0$ (in the two-color spectroscopy case). The color coding in panel \textbf{d} corresponds to that in panel \textbf{c}.}
  \label{fig5}
\end{figure}

Because the time shift of the maximum harmonic intensity induced by saturation is highly sensitive to the electron population, it can serve as an effective probe for tracking electron population dynamics. To investigate this, we analyze the electron dynamics associated with H4 generation through the interband process by integrating the conduction band population over the momentum-space region where the bandgap energy corresponds to the H4 photon energy (see Fig.~\ref{fig5}a).

Under excitation by a 2~$\mu$m pulse with a peak intensity of 70\,GW\,cm$^{-2}$, the electron population contributing to H4 exhibits strong temporal oscillations (see Fig.~\ref{fig5}b). We define $t_c$ as the time at which a quarter of the electrons in this momentum-space region have transitioned to the conduction band. As shown in Fig.~\ref{fig5}c, the absolute value of $t_c$ increases with laser intensity, indicating an intensity-dependent excitation process. Furthermore, a strong correlation is observed between $t_c$ (in the single-color 2 $\mu$m excitation case) and $\mu_0$ (emission time shift corresponding to the maximum H4 intensity in the two-color excitation case). The correspondence can be captured by a second-order polynomial fitting function, $t_c = 0.027\mu_0^2 + 0.63\mu_0 - 16.71$, with an $R^2$ value of 0.99, indicating very strong correlation between these two physical observables. Unlike transient absorption spectroscopy, which mostly probes post-excitation relaxation dynamics, our two-color harmonic spectroscopy directly tracks the coherent, ultrafast carrier dynamics while they are driven by the strong optical field. This demonstrates an all-optical method for probing carrier population dynamics in Dirac materials and semi-metallic systems under strong-field excitation.

\section*{Conclusion}\label{sec13}  

In conclusion, our joint experimental and theoretical study has established an NPHG-based all-optical approach to observe and quantify ultrafast carrier saturation effects and carrier excitation dynamics in solids, which also yields absolute timing information. With respect to conventional bulk materials, such as ZnO, we found that the Dirac semimetal HOPG shows strong conduction band population saturation due to its gapless nature even under moderate laser intensities ($\sim$$10^{10}\,\mathrm{W\,cm}^{-2}$), leading to half-filling in the vicinity of the Dirac point. We identified this effect as responsible for the observed $\sim 17.5$ fs time shifts and the suppression of the harmonic emission in the later part of the driving laser pulse through state blocking.

Our study demonstrates that gapless materials are subject to saturation effects and state blocking, leaving behind a sizable conduction band population and rendering harmonic generation processes nonparametric in nature. Petahertz optoelectronics, where optical manipulation of the material is performed on attosecond time scales, relies mainly on virtual charge carrier excitations and their reversibility~\cite{Sommer2016,Boulakee2022,Borsch2023}. In order to realize petahertz optoelectronics with full reversibility in gapless materials, pulse duration and peak intensity need to be limited such that state blocking is avoided. Even if this cannot be achieved, gapless materials can nevertheless serve as an ultrafast optoelectronic switch where the presence or the absence of a pre-excitation pulse affects the efficiency of nonlinear processes through state blocking on the significantly longer time scale of spontaneous emission. The nonparametric nature of HHG in gapless materials even at moderate intensities may also be helpful for generating upconverted light with nonclassical photon statistics due to a saturation or depletion effect, as recently predicted for gas-phase HHG~\cite{Stammer2024}.

\section*{Methods}

\subsection*{Experimental setup}

In the experiment, the fundamental ($\omega_0$) pulse at 1980 nm is generated from a phase-stable optical parametric chirped pulse amplification (OPCPA) laser system operating at a repetition rate of 1\,MHz (Class 5 Photonics White Dwarf~\cite{Braatz2021}). The pulse duration of the $\omega_0$ field measured by frequency-resolved optical gating (FROG) yields a full-width-at-half-maximum (FWHM) duration of approximately 60 fs ($\sim9$ optical cycles). The second harmonic ($2\omega_0$) pulse was generated by focusing the fundamental beam into a $\beta$-barium borate (BBO) crystal under optimum phase-matching conditions. By adjusting the $\omega_0$ power, we maintained the intensity ratio between the $2\omega_0$ and $\omega_0$ pulses at $I_{2\omega_0}/I_{\omega_0} = 3\%$. For HHG in HOPG, the $\omega_0$ pulse energy was set between 1–5\,nJ, while for ZnO, a higher energy of approximately 30\,nJ was used.

The two-color interferometer setup was implemented by first separating the $\omega_0$ and $2\omega_0$ beams using a dichroic beam splitter. The relative delay between the two pulses was controlled via a piezo delay stage placed in the $2\omega_0$ beam path. The two beams were then recombined by reflecting them off two mirrors back into the same dichroic beam splitter, with a slight lateral offset angle relative to the incoming paths, and subsequently directed toward the sample for harmonic generation. A parabolic mirror with $f = 25$\,mm focuses the light onto the HOPG sample (TipsNano) at an angle of incidence of 14 degrees.  The beam size at the focus was measured using a knife-edge method, yielding an effective $1/e^2$ radius of $\approx 8.35\ \mu\mathrm{m}$ (see discussion and Fig.~S1 in the Supplementary Information). We collect the harmonic light with a lens and send it to a Si-based spectrometer.

During the two-color delay scan, the piezo stage exhibited a minimal drift of less than 1\,fs, which does not affect the accuracy of the delay-dependent measurements. Furthermore, to actively monitor and correct for any such drift, a portion of the combined $\omega_0$–$2\omega_0$ beam was directed into an additional BBO crystal to generate a reference SH signal. The resulting modulation arises from the interference between the newly generated SH signal from the $\omega_0$ beam and the original $2\omega_0$ pulse. This delay-dependent modulation serves as a timing reference, allowing for post-correction of any small delay drifts in the piezo stage.

The thickness of the reference ZnO crystal (FarView, $\left<0001\right>$) is 0.5 mm. Since the harmonics are measured in a reflection geometry on the crystal surface, nonlinear propagation effects associated with the crystal thickness do not influence our results. 

\subsection*{Semiconductor Bloch equations calculations}

To describe the interaction of the two-color field and HOPG in the Dirac cone, we use the following one-dimensional semiconductor Bloch equations\cite{Koch2008SBE,luu2016high} (SBEs) centered around the Dirac cone:
\begin{equation}\label{SBE3_P}
\begin{aligned}
i \frac{\partial}{\partial t} p_k(t) &= \left[ \epsilon_k^{c} - \epsilon_k^{v} - i \frac{1}{T_2} + i E(t) \nabla_k \right] p_k(t) 
 - \left( 1 - f_k^{c}(t) - f_k^{v}(t) \right) d^{c \nu}_k E(t), \\
\frac{\partial}{\partial t} f_k^{\nu}(t) &= -2 \text{Im} \left[ d_k^{c v} E(t) \left( p_k^{\nu c}(t) \right)^* \right] + \left [ -\frac{1}{T_1}+ E(t) \nabla_k\right ] f_k^{\nu}(t), \\
\frac{\partial}{\partial t} f_k^{c}(t) &= -2 \text{Im} \left[ d_k^{\nu c} E(t) \left( p_k^{c \nu}(t) \right)^* \right] + \left [ -\frac{1}{T_1}+ E(t) \nabla_k\right ] f_k^{c}(t). 
\end{aligned}
\end{equation}
Here, $p_k(t)$ represents a dimensionless polarization depending on time $t$ and momentum $k$, induced by the time-dependent electric field $E(t)$. The indices $v$ and $c$ correspond to the valence band and conduction band, respectively. $f_k^{c,\nu}(t)$ denotes the population of electrons (holes) in the conduction (valence) band. Here, we consider linear energy-momentum dispersion relation around Dirac cone with Fermi velocity $v_F=1 \times 10^6\,\mathrm{m\,s}^{-1}$ describing a continuum ground state Dirac Hamiltonian. This approximation should be valid in our regime where the laser-driven electrons only move near the Dirac cone. The transition dipole moments (TDMs) are calculated as $d_k^{\nu c}=1/(2|k|)$ ~\cite{al2014high} (see Fig.~S3 in the Supplementary Information). We set the phenomenological interband polarization decoherence time to one optical cycle of the 2 $\mu$m driving laser field $T_2=6.6$ fs, and the relaxation time for intraband current to $T_1=13\,\mathrm{ps}$. This consideration is valid since the relaxation is generally long and ranges from few hundred femtosecond to picosecond in graphene and is less sensitive to the results~\cite{sato2021high,liu2018driving}. Band structure and TDM for ZnO crystal along the $\Gamma$–M direction are shown in Fig.~S4 of the Supplementary Information (for details see the Supplementary Information). The calculations are performed with dense $k$-grid of size 200, using a time step of 2.4 as and according to the experimental parameters. The  yield of the harmonics for the two-color case are calculated with two-color delay steps of $\sim$ 0.33 fs.     

The HHG spectrum is computed as~\cite{Koch2008SBE}
\begin{equation}
I_{\text{HHG}}(\omega) \sim \left| \int_{-\infty}^{\infty} [J_{\text{inter}}(t) + J_{\text{intra}}(t)] e^{i\omega t} dt \right|^2,
\end{equation}
where 
\begin{equation}
J_{\text{inter}}(t) = \frac{d}{dt} \int d_k^{\nu c} \cdot (p_k (t))^* dk + \text{c.c.}
\end{equation}
and
\begin{equation}
J_{\text{intra}}(t) = \sum_{\lambda=c,\nu} \int \nu_k^{\lambda}  f_k^{\lambda}(t) dk
\end{equation}
denote the interband and intraband contributions to the harmonic yield, respectively. $\nu_k^{\lambda} = \nabla_k \epsilon^{\lambda}_k$ represents the group velocity of the electrons and holes in the corresponding bands.

\subsection*{Analytical form of population dynamics}

For a theoretical analysis of saturation and electron population dynamics, we begin by considering the \textit{non-depleted} excitation rate of electrons from the valence band to the conduction band. The excitation rate at a given crystal momentum \( k \) and time \( t \) can be expressed as

\begin{align}
w_0(k, t) &\sim \frac{2\pi}{\hbar} \left| \langle \psi_c | \mathbf{E}(t) \cdot \hat{\mathrm{d}} | \psi_v \rangle \right|^2 
\approx 2\pi \left| \mathbf{E}(t) \cdot \mathbf{d}_{cv}(k+A(t)) \right|^2.
\end{align}

The total excitation rate integrated over the Brillouin zone is then given by
\begin{align}
w(t) = \int w_0(k, t) \, dk = 2\pi E^2(t) \int |\mathbf{d}_{cv}(k+A(t))|^2 \, dk. \label{nondeple_excit_rate}
\end{align}

Let \( P(t) \) denote the total conduction band electron population at time \( t \). The evolution of this population in discrete time steps \( t_n \rightarrow t_{n+1} \) can be described by

\begin{align}
& P(t_{n+1}) = P(t_n) + (1 - P(t_n)) w(t_n)\mathrm{d}t - P(t_n) w(t_n)\mathrm{d}t, \\
& \frac{P(t_{n+1}) - P(t_n)}{1 - 2P(t_n)} = w(t_n)\mathrm{d}t. \label{Popul1}
\end{align}

Transitioning to the continuous-time limit, we obtain the differential form
\begin{align}
&\mathrm{d}P(t)/(1 - 2P(t)) = w(t)\mathrm{d}t,\\
&\mathrm{d}\ln(1 - 2P(t)) = -2w(t)\mathrm{d}t.
\end{align}

Integrating yields the electron population as a function of time,
\begin{equation}
P(t) = \frac{1 - e^{- \int_0^t 2w(\tau)\, d\tau}}{2}.
\end{equation}

The actual (depleted) instantaneous excitation rate is given by

\begin{equation}
w_\mathrm{depletion}(t) =dP(t)/dt= e^{- \int_0^t 2w(\tau)\, d\tau} \cdot w(t). \label{analy_wt}
\end{equation}

For interband-driven harmonic generation, the emitted harmonic intensity is proportional to the instantaneous excitation rate, $I_\mathrm{HH}(t) \sim w_\mathrm{depletion}(t) $.

\section*{Data availability}

The data that support the plots within this paper and other findings of this study are available from the authors upon request.

\section*{Code availability}

The codes that support the findings of this study are available from the authors upon request.

\section*{Author contributions}

Z.~C. and C.~G.~contributed equally to this work.

M.~K.~and M.~F.~C.~conceived and supervised the project, joined by O.~N.~in the later stage of the project. Z.~C.~built the experimental setup and performed the measurements. C.~G.~performed initial simulations. Z.~C.~carried out the theory calculations and C.~G.~and E.~U.~and O.~N.~performed independent supplemental calculations. Z.~C.~wrote the initial manuscript. All authors contributed to the interpretation of the results and to the preparation of the final manuscript.

\section*{Corresponding authors}

Correspondence to \href{mailto:krueger@technion.ac.il}{Michael Kr\"uger} or \href{mailto:marcelo.ciappina@gtiit.edu.cn}{Marcelo F.~Ciappina} or \href{mailto:ofern@technion.ac.il}{Ofer Neufeld}.

\begin{acknowledgments}
We thank M.~Ivanov, M.~Segev, Q.~Yan and Y.~Wu for insightful discussions, U.~Leonhardt and L.~M.~Procopio for providing specialized equipment, and X. Duan for providing the monolayer WS$_2$ sample.  Z.~C., C.~G., I.~N., D.~K., M.~F.~C.~and M.~K.~acknowledge the Guangdong Technion -- Israel Institute of Technology (GTIIT) and Technion Seed Grant Program for enabling their collaboration and joint research. Z.~C., I.~N., D.~K., E.~U., O.~N.~and M.~K.~thank the Helen Diller Quantum Center  and the Russell Berrie Nanotechnology Institute at the Technion for partial financial support. M.~F.~C.~acknowledges support by the National Key Research and Development Program of China (Grant No.~2023YFA1407100), Guangdong Province Science and Technology Major Project (Future functional materials under extreme conditions - 2021B0301030005), the Guangdong Natural Science Foundation (General Program project No. 2023A1515010871) and the National Natural Science Foundation of China (Grant No. 12574092).
\end{acknowledgments}

\section*{Ethics declarations}

The authors declare no competing interests.




\end{document}


\title{Supplementary Information:
Two-color harmonic spectroscopy of ultrafast Dirac electron dynamics}

\author{Zhaopin Chen}
\altaffiliation{These authors contributed equally to this work.}
\affiliation{Department of Physics, Technion -- Israel Institute of Technology, Haifa 3200003, Israel}
\affiliation{Solid State Institute, Technion -- Israel Institute of Technology, Haifa 3200003, Israel}
\affiliation{The Helen Diller Quantum Center, Technion -- Israel Institute of Technology, Haifa 3200003, Israel}

\author{Camilo Granados}
\altaffiliation{These authors contributed equally to this work.}
\affiliation{Department of Physics, Technion -- Israel Institute of Technology, Haifa 3200003, Israel}
\affiliation{Department of Physics, Guangdong Technion -- Israel Institute of Technology, Shantou 515063, Guangdong, China}
\affiliation{Guangdong Provincial Key Laboratory of Materials and Technologies for Energy Conversion, Guangdong Technion -- Israel Institute of Technology, Shantou 515063, Guangdong, China}
\affiliation{Eastern Institute of Technology, Ningbo 315200, China}

\author{Eyal Uzner}
\affiliation{Department of Physics, Technion -- Israel Institute of Technology, Haifa 3200003, Israel}
\affiliation{Schulich Faculty of Chemistry, Technion -- Israel Institute of Technology, Haifa 3200003, Israel}

\author{Ido Nisim}
\affiliation{Department of Physics, Technion -- Israel Institute of Technology, Haifa 3200003, Israel}
\affiliation{Solid State Institute, Technion -- Israel Institute of Technology, Haifa 3200003, Israel}
\affiliation{The Helen Diller Quantum Center, Technion -- Israel Institute of Technology, Haifa 3200003, Israel}

\author{Daniel Kroeger}
\affiliation{The Norman Seiden Multidisciplinary Graduate Program in Nanoscience and Nanotechnology, Technion -- Israel Institute of Technology, Haifa 3200003, Israel}
\affiliation{Solid State Institute, Technion -- Israel Institute of Technology, Haifa 3200003, Israel}
\affiliation{The Helen Diller Quantum Center, Technion -- Israel Institute of Technology, Haifa 3200003, Israel}

\author{Ofer Neufeld}
\affiliation{Schulich Faculty of Chemistry, Technion -- Israel Institute of Technology, Haifa 3200003, Israel}

\author{Marcelo F. Ciappina}
\affiliation{Department of Physics, Technion -- Israel Institute of Technology, Haifa 3200003, Israel}
\affiliation{Department of Physics, Guangdong Technion -- Israel Institute of Technology, Shantou 515063, Guangdong, China}
\affiliation{Guangdong Provincial Key Laboratory of Materials and Technologies for Energy Conversion, Guangdong Technion -- Israel Institute of Technology, Shantou 515063, Guangdong, China}

\author{Michael Krüger}
\affiliation{Department of Physics, Technion -- Israel Institute of Technology, Haifa 3200003, Israel}
\affiliation{Solid State Institute, Technion -- Israel Institute of Technology, Haifa 3200003, Israel}
\affiliation{The Helen Diller Quantum Center, Technion -- Israel Institute of Technology, Haifa 3200003, Israel}


\keywords{}

\maketitle

\newpage

\section*{Supplementary text}\label{sec0}

\subsection*{Knife-edge measurement of the focus}

\begin{figure}[htb!]
\centering
\includegraphics[width=0.9\textwidth]{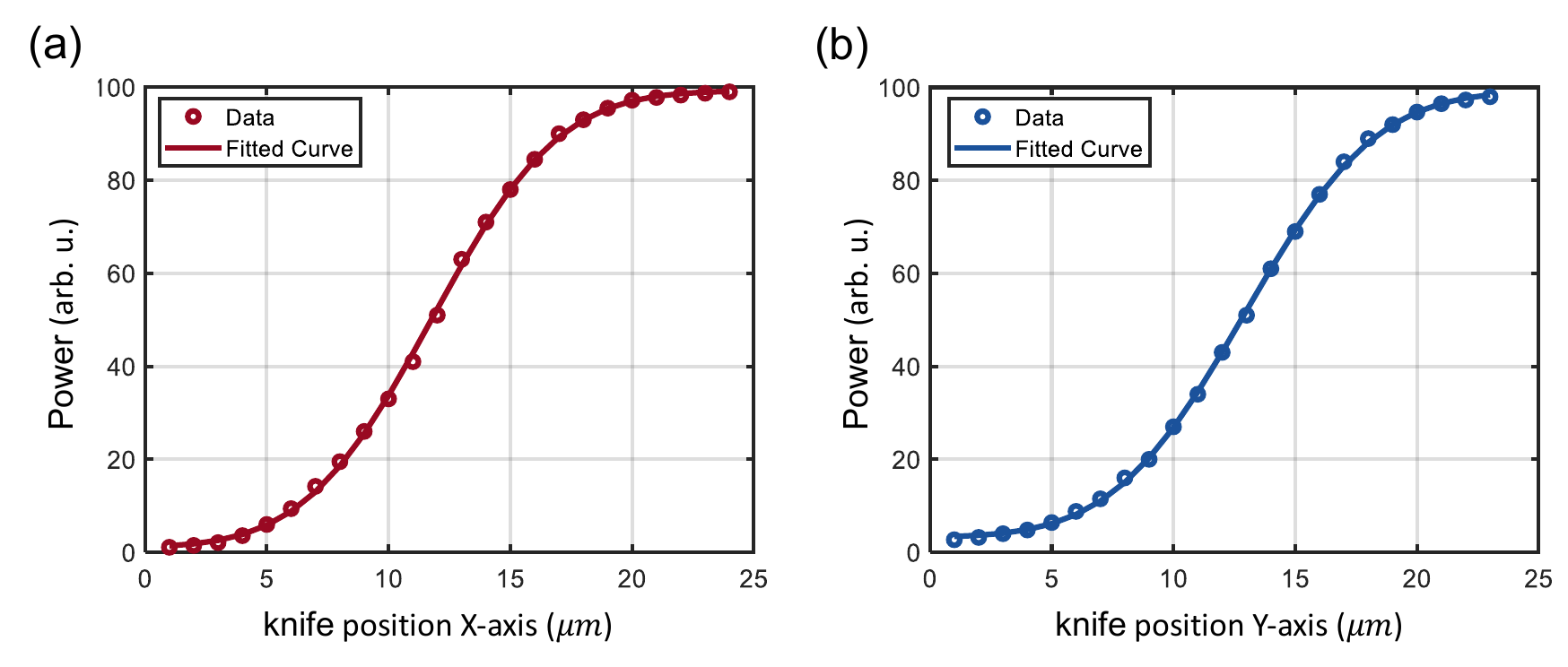} 
\caption{Knife-edge measurement of the beam size at the focus. The transmitted power is fitted using the function
$P_t(h_i) = P_{\text{offset}} + \frac{P}{2} , \mathrm{erfc} \left[ \frac{h_i - h_0}{w / \sqrt{2}} \right]$,
where $w$ denotes the $1/e^2$ beam radius. From the fit, we obtain $a_x \approx 8.2\ \mu\mathrm{m}$ along the X-axis and $a_y \approx 8.71\ \mu\mathrm{m}$ along the Y-axis. The effective beam radius at the focus is therefore $r = \sqrt{a_x a_y} \approx 8.35\ \mu\mathrm{m}$.}
\label{beam_size}
\end{figure}

We characterize the driving laser beam size at the focus using a knife-edge measurement after a parabolic mirror with a focal length of $f = 25\,\mathrm{mm}$. As the knife-edge gradually unblocks the beam, the transmitted power increases accordingly, as shown in Fig.~\ref{beam_size}. Panels (a) and (b) show measurements along the X and Y axes, respectively. The experimental data are fitted using the function: $P_t(h_i) = P_{\text{offset}} + \frac{P}{2} , \mathrm{erfc} \left[ \frac{h_i - h_0}{w / \sqrt{2}} \right]$, where $w$ is the $1/e^2$ beam radius. From the fit, we obtain $a_x \approx 8.2\ \mu\mathrm{m}$ along the X-axis and $a_y \approx 8.71\ \mu\mathrm{m}$ along the Y-axis. The effective beam radius at the focus is then calculated as $r = \sqrt{a_x a_y} \approx 8.35\ \mu\mathrm{m}$.

\subsection*{Measured harmonic spectra as functions of $\omega_0-2\omega_0$ delay}

\begin{figure}[htb!]
\centering
\includegraphics[width=1\textwidth]{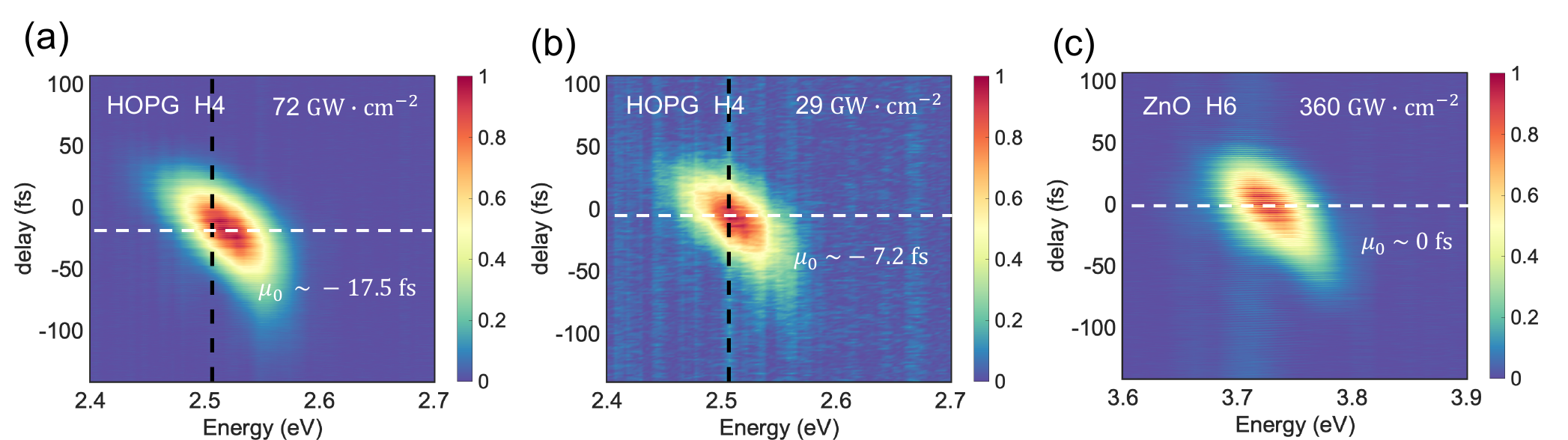} 
\caption{Experimentally measured harmonic spectra as functions of the two-color delay, corresponding to Fig.~2 in the main text. (a) and (b) Fourth harmonic (H4) from HOPG at peak intensities of 72 and 29 $\mathrm{GW\ cm}^{-2}$, respectively. (c) sixth harmonic (H6) from ZnO at a peak intensity of 430 $\mathrm{GW\ cm}^{-2}$. The white dashed lines indicate the delay times corresponding to the maximum harmonic intensities. The black dashed lines mark the energy of $4\omega_0 = 2.50$ eV, serving as a reference to highlight the energy shift. } \label{Spect_2delay}
\end{figure}

Experimentally measured harmonic spectra as functions of the two-color delay are presented in Fig.~\ref{Spect_2delay}, corresponding to Fig.~2 in the main text, but without integrating over specific harmonic energies. Fig.~\ref{Spect_2delay}(a) and (b) show the fourth harmonic from HOPG at peak intensities of 72 and 29 $\mathrm{GW\ cm}^{-2}$, respectively, while Fig.~\ref{Spect_2delay}c displays the sixth harmonic (H6) from ZnO at a peak intensity of 430 $\mathrm{GW\ cm}^{-2}$ as a reference. For HOPG, it is evident that the time shift corresponding to the maximum harmonic intensity moves to earlier times at higher laser intensities, indicating stronger saturation effects. In contrast, the lower intensity case exhibits much smaller saturation. For ZnO, with its much larger bandgap ($\sim$$3.3$ eV), there is no saturation even at 430 $\mathrm{GW\ cm}^{-2}$. Therefore, the maximum harmonic signal consistently appears at zero delay, corresponding to optimal temporal overlap of the fundamental and second harmonic pulses. 
In addition, for HOPG the fourth harmonic spectrum exhibits a clear blueshift at negative delay and a redshift at positive delay with respect to $4\omega_0$ energy. Increasing the pump intensity further enhances the overall blueshift of the spectrum. 

\subsection*{Band structure and transition dipole moment of HOPG and ZnO in 1D SBEs}

\begin{figure}[htb!]
\centering
\includegraphics[width=0.9\textwidth]{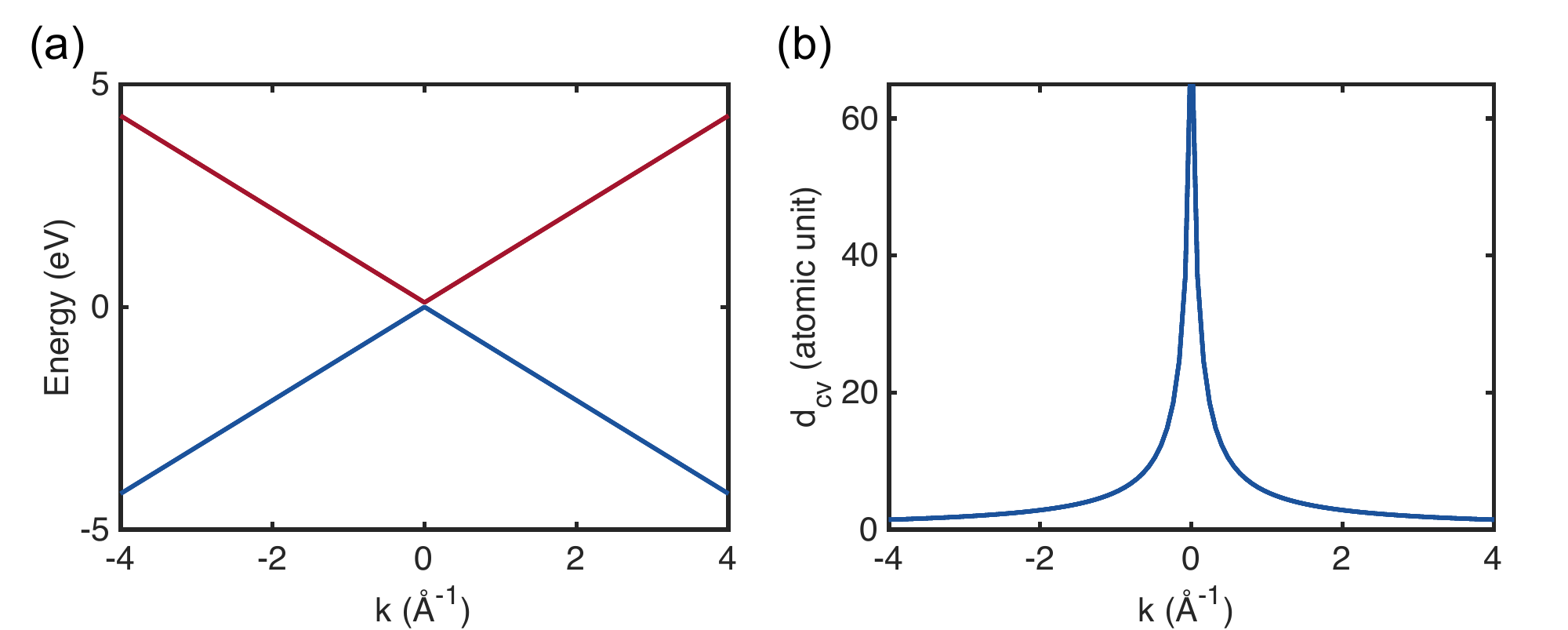} 
\caption{(a) One-dimensional band structure and (b) transition dipole moment of graphene-like HOPG around Dirac cone used in the SBE calculations. }\label{HOPG_band}
\end{figure}

In our one-dimensional semiconductor Bloch equation (SBE) simulations, we model the electronic structure of HOPG by considering only the region near the Dirac cone, with a Fermi velocity of $V_F = 10^6\ \mathrm{m\,s^{-1}}$. Within this linear dispersion regime, the bandgap is given by $E_g(k) = E_c(k) - E_v(k) = 2V_F|k|$, and the interband transition dipole moment (TDM) is defined as $d_k^{vc} = 1/(2|k|)$, following Ref.\cite{al2014high} (see Fig.\ref{HOPG_band}). To avoid divergence at $k = 0$, we regularize the TDM by introducing a small cutoff, yielding $d_k^{vc} = 1/(2|k| + \Delta k)$ in our numerical implementation, where $\Delta k$ corresponds to the discretization step in $k$-space.

\begin{figure}[htb!]
\centering
\includegraphics[width=0.9\textwidth]{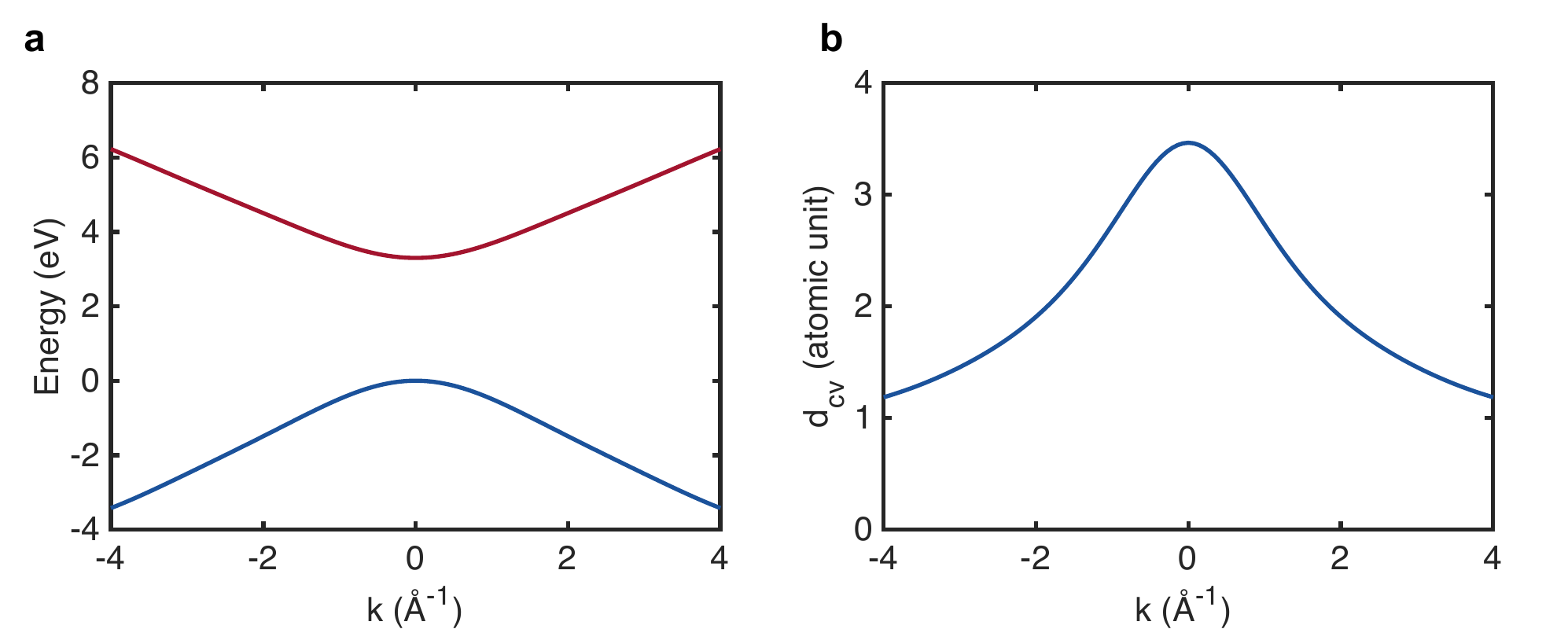} 
\caption{(a) One-dimensional band structure and (b) transition dipole moment of ZnO along $\Gamma-M$ orientation used in the SBE calculations.}\label{ZnO_band}
\end{figure}

For ZnO, we use the band structure and transition dipole moment along the $\Gamma$–M direction, taken from references~\cite{vampa2015b,yu2016dependence}, as shown in Fig.~\ref{ZnO_band}. Here, the ZnO TDM is defined as $d_k^{vc} = d_0E_g(0)/E_g(k)$ with $E_g(k)=E_c(k)-E_v(k)$ and $d_0=3.46$ (a.u.).

\subsection*{Interband and intraband harmonics spectra in HOPG}

\begin{figure}[htb!]
\centering
\includegraphics[width=0.6\textwidth]{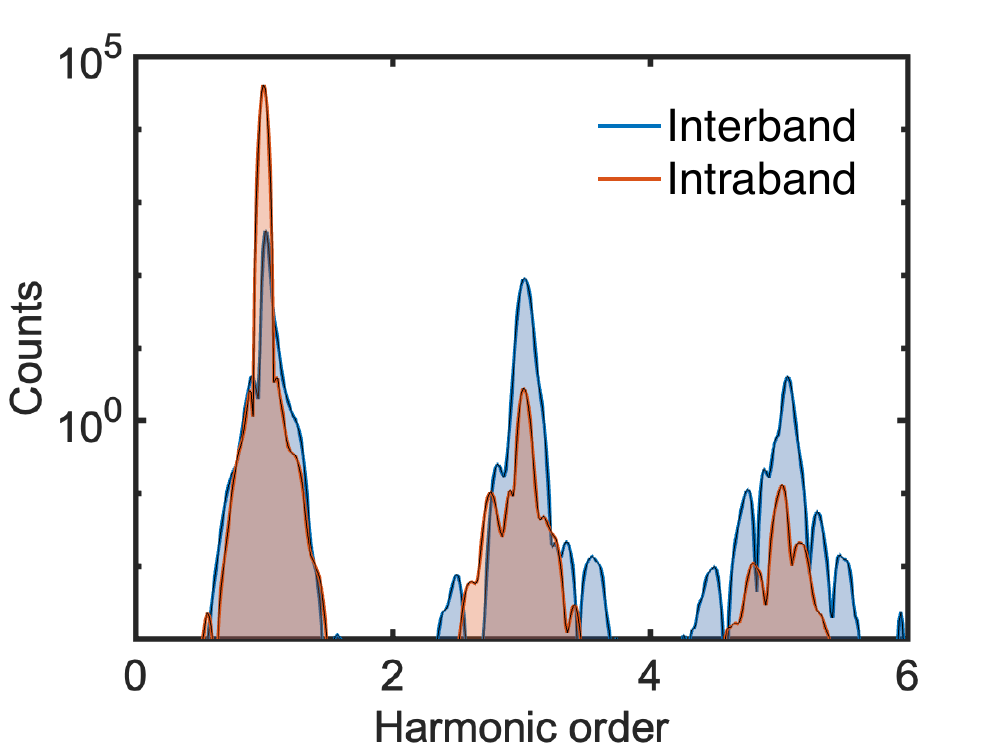} 
\caption{Harmonic spectra simulated using the semiconductor Bloch equations (Eqs. (1) in the main text), driven solely by a 1980 $\mathrm{n m}$ laser pulse at a peak intensity of 70 $\mathrm{GW\ cm^{-2}}$ in HOPG. The interband harmonics clearly dominate, with intensities approximately two orders of magnitude higher than the intraband harmonics. }\label{Inter_intra}
\end{figure}

Figure~\ref{Inter_intra} compares the harmonic spectral intensity contributions from interband and intraband currents in HOPG, driven by a 2\,$\mu m$ pulse with a peak intensity of $70\,\mathrm{GW\ cm}^{-2}$. The results clearly show that the harmonic yield from the interband current is approximately two orders of magnitude higher than that from the intraband current.

\subsection*{SBEs simulated harmonic spectra as functions of $\omega_0-2\omega_0$ delay}

\begin{figure}[htb!]
\centering
\includegraphics[width=1\textwidth]{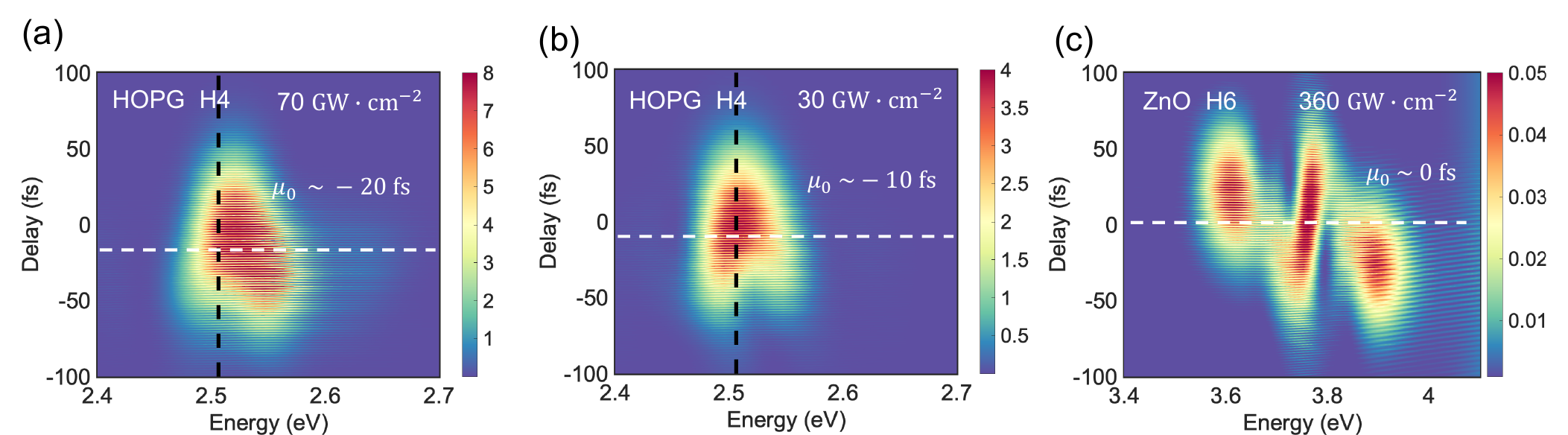} 
\caption{SBEs simulated harmonic spectra as functions of the two-color delay, corresponding to Fig.~2 in the main text: (a, b) fourth harmonic (H4) from HOPG at peak intensities of ~70 and ~30 $\mathrm{GW\ cm}^{-2}$, respectively; (c) sixth harmonic (H6) from ZnO at a peak intensity of 430 $\mathrm{GW\ cm}^{-2}$. The white dashed lines indicate the delay times corresponding to the maximum harmonic intensities. The black dashed lines mark the energy of $4\omega_0 = 2.50$ eV, serving as a reference to highlight the energy shift. } \label{Spect_2delay_num}
\end{figure}

Simulated harmonic spectra as functions of the two-color delay, based on the 1D SBEs, are shown in Fig.~\ref{Spect_2delay_num}. These correspond to Fig.~2(c) and (d) in the main text, but without integration over specific harmonic energies. Fig.~\ref{Spect_2delay_num}(a) and (b)  display the fourth harmonic from HOPG at peak intensities of 70 and 30 $\mathrm{GW\ cm}^{-2}$, respectively. Figs.~\ref{Spect_2delay_num}(c)  shows the sixth harmonic from ZnO at a peak intensity of 430 $\mathrm{GW\ cm}^{-2}$ as a reference.

Compared to the experimental results, the 1D SBE simulations accurately capture the emission time shift corresponding to the maximum harmonic emission ($\mu_0$) in HOPG. The simulations also reveal a redshift in the central harmonic energy as the delay increases, consistent with our experimental observations.

For ZnO, the maximum H6 intensity remains centered around $\mu_0 = 0$, indicating negligible saturation effects due to the large bandgap. In addition to the redshift, the H6 spectrum exhibits a clear splitting at larger delays. This feature may arise from interference between different quantum trajectories, such as short and long trajectories, which are likely smeared out in the experiment due to spatial volume averaging.

Since the harmonics generated in HOPG are above the (zero) bandgap and predominantly originate from interband currents, it is more appropriate to compare them with above-bandgap harmonics in ZnO. Therefore, we use the sixth harmonic from ZnO as a reference, rather than H4, which lies below the bandgap under our two-color excitation conditions.

\begin{figure}[htb!]
\centering
\includegraphics[width=0.6\textwidth]{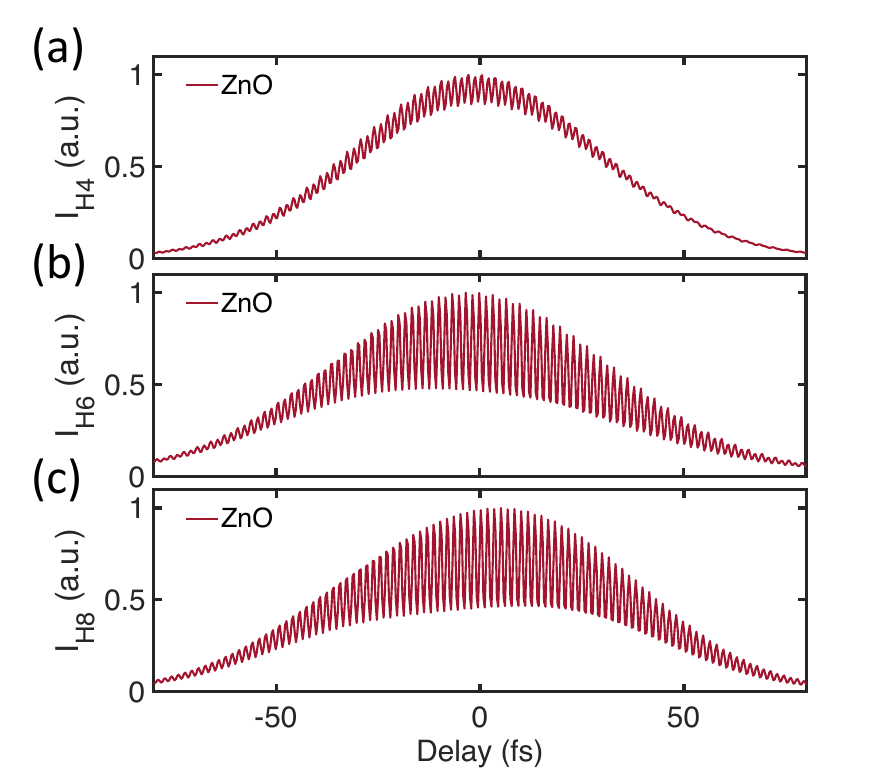} 
\caption{Simulated harmonic intensities in ZnO as functions of the $\omega_0$–$2\omega_0$ time delay for (a) the 4th harmonic (H4), (b) the 6th harmonic (H6), and (c) the 8th harmonic (H8), respectively.}\label{Delay_ZnO}
\end{figure}

Nevertheless, we also compare the simulated harmonic intensities of H4, H6, and H8 in ZnO as functions of the two-color delay. Due to the absence of saturation effects, the peak intensities for all three harmonics are centered around zero delay. An exception is H8, which exhibits a slight shift (less than 5 fs) toward positive delay, see Fig.~\ref{Delay_ZnO}.

\subsection*{Time-frequency analysis for two-color HHG in HOPG}

\begin{figure}[htb!]
\centering
\includegraphics[width=0.6\textwidth]{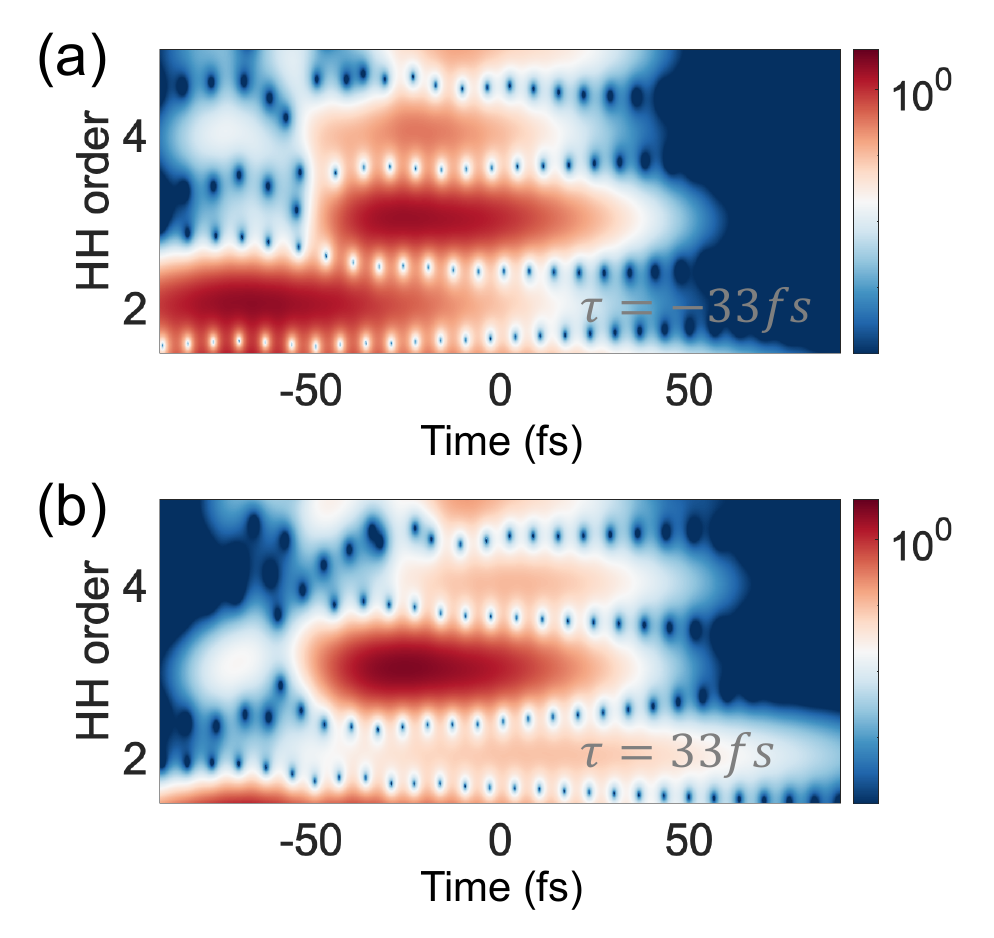} 
\caption{Gabor transform of the harmonic emission generated by $\omega_0$–$2\omega_0$ two-color pulses at different delay times: (a) $\tau = -33,\mathrm{fs}$ and (b) $\tau = 33,\mathrm{fs}$. The driving laser intensity is fixed at 70 $\mathrm{GW\ cm}^{-2}$. }\label{2color_Gabor}
\end{figure}

To investigate the relationship between single-color and two-color driven HHG in HOPG, we apply a Gabor transform to analyze the time–frequency distribution of the harmonic emission driven by the two-color field at two different delays, as shown in Fig.~\ref{2color_Gabor}. At a negative delay of $\tau = -33\,\mathrm{fs}$, the fourth harmonic exhibits strong emission, primarily localized at negative propagation times that correspond to the delay. In contrast, for a positive delay of $\tau = 33\,\mathrm{fs}$, the H4 intensity is significantly weaker and shifts to positive propagation times. These results suggest that the $2\omega_0$ field acts as a probe, sampling the harmonic emission that originates primarily from the single-color ($\omega_0$) excitation.

\subsection*{High harmonics generated from ZnO with an artificially adjusted band gap }

\begin{figure}[htb!]
\centering
\includegraphics[width=0.9\textwidth]{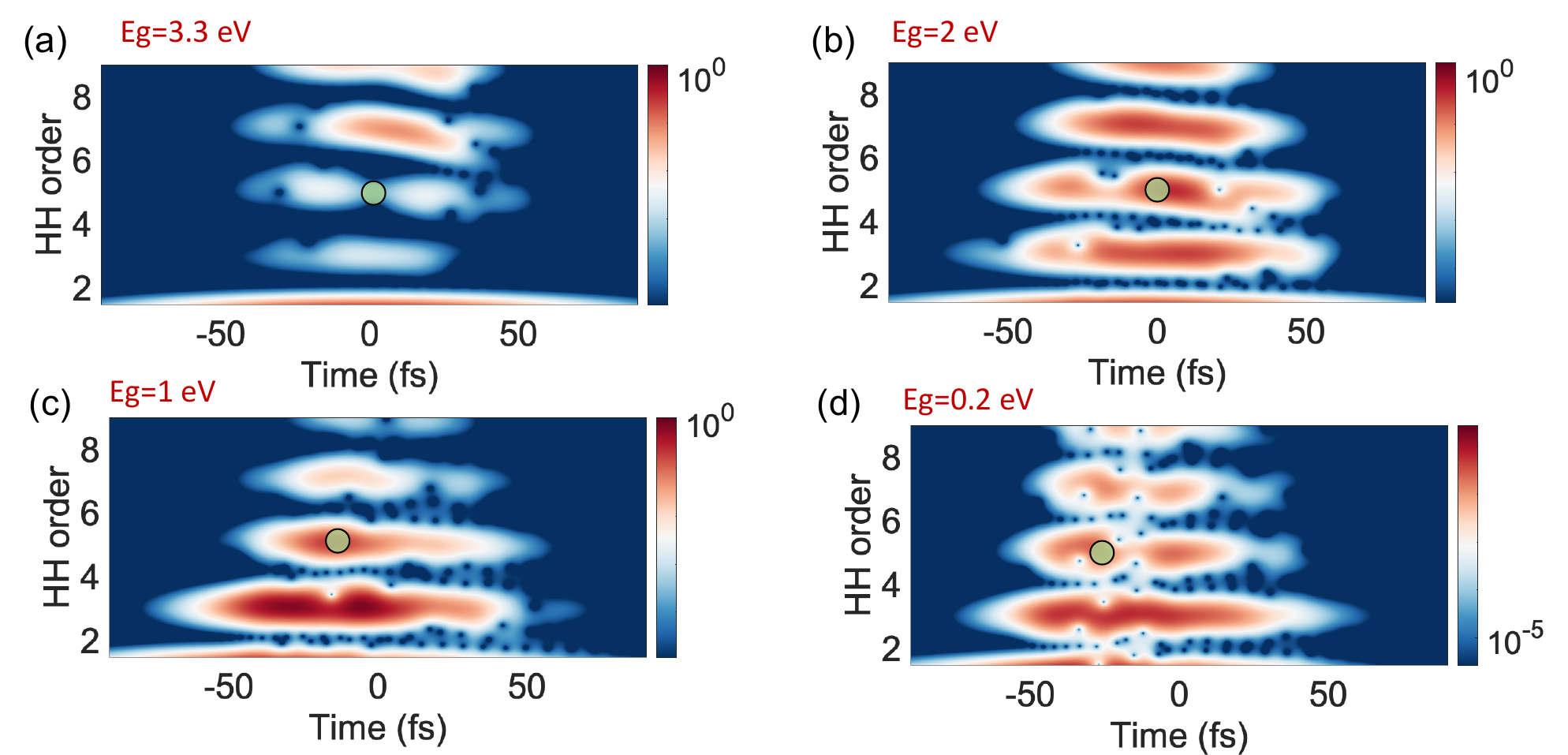} 
\caption{Time-frequency analysis of interband harmonics using a Gabor transformation at a 1980\,nm peak intensity of $I = 430\,\mathrm{GW\ cm}^{-2}$. The maximum intensity of the harmonics is shown for ZnO with manually adjusted bandgaps: (a) $E_g = 3.3\,\mathrm{eV}$, (b) $E_g = 2.0\,\mathrm{eV}$, (c) $E_g = 1.0\,\mathrm{eV}$, and (d) $E_g = 0.2\,\mathrm{eV}$. A clear shift to negative emission times is observed as the bandgap decreases, indicating earlier saturation of the valence band due to more easily driven interband transitions in narrower-gap materials. }\label{ZnO_Gabor}
\end{figure}

The saturation effect observed in HOPG primarily arises from its gapless band structure, which enables significant electron transitions from the valence to the conduction band. Consequently, similar behavior can be expected in other materials with zero or small band gaps. To explore the influence of the band gap on saturation-induced high-harmonic generation, we artificially reduce the band gap in ZnO and analyze the harmonic emission dynamics using time-frequency (Gabor) analysis. As shown in Fig.~\ref{ZnO_Gabor}, a clear shift of the harmonic emission toward earlier times is observed as the band gap decreases. Specifically, the emission time corresponding to the maximum intensity of the fifth harmonic (H5) advances with decreasing band gap, indicating a direct correlation between the band gap size and the saturation-induced temporal shift in harmonic emission.

\subsection*{Numerical simulation by 2D Semiconductor Bloch equations}

To further explore the validity of the simpler 1D SBE numerical approach, we employed a more general 2D SBE simulation performed in the length gauge and in the Houston basis \cite{Yue2022}. The numerical approach follows that described in detail in ref. \cite{kim2025quantum}, where the equations of motion and procedures can be found. The graphene system is considered with a two-band tight binding Hamiltonian with up to 14 nearest-neighbor hopping terms, which are fitted to density functional theory (DFT) simulations of the ground state bands (following the procedures in Refs. \cite{neufeld2023band} and \cite{Galler2023}). This provides high-accuracy bands around the Dirac cones at $K$ and $K$'. Simulations were performed with a dense 360$\times$360 $k$-grid, using a time step of 4.8 as and following the experimental conditions. We employed a $T_2$ value of 6.6 fs, similar to the 1D model, and also tested longer $T_2$ of 25 fs. The harmonic yield was integrated around the peak without any filtration and calculated vs. two-color delay in steps of $\sim1.5$ fs. 

\begin{figure}[htb!]
\centering
\includegraphics[width=0.9\textwidth]{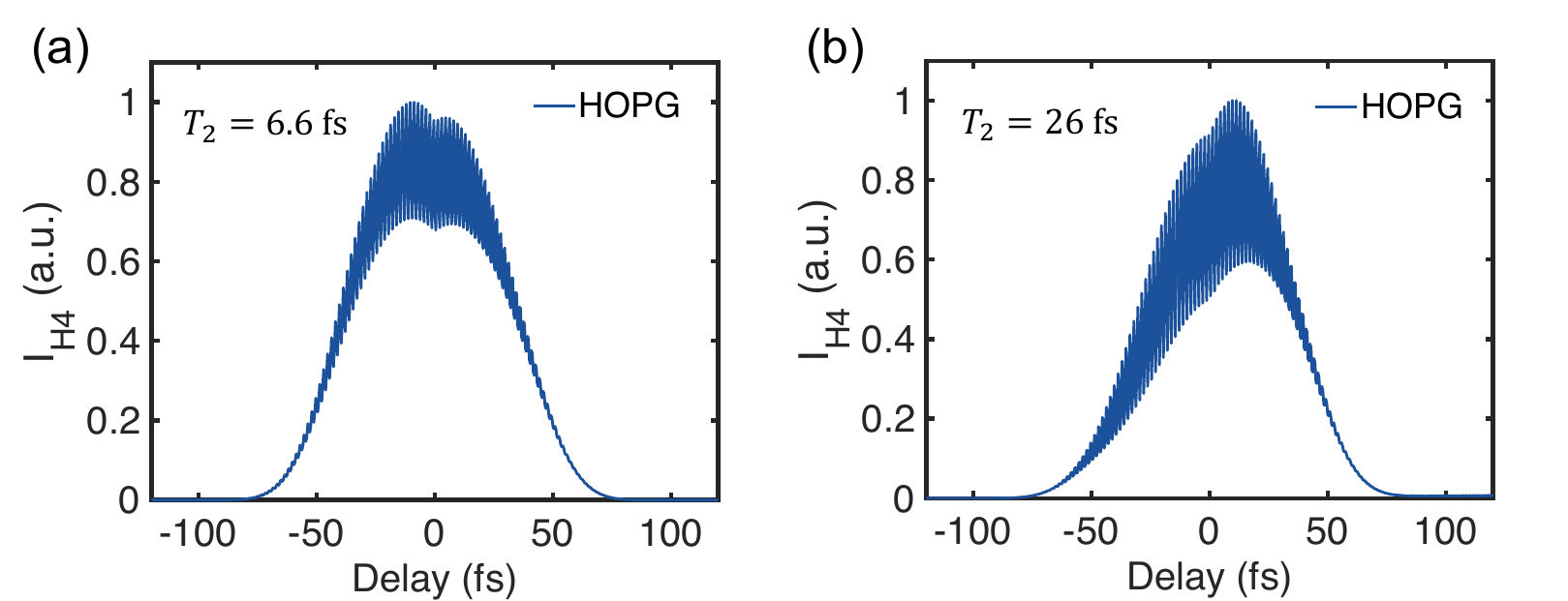} 
\caption{Fourth harmonic intensity in HOPG as a function of the $\omega_0$–$2\omega_0$ time delay, simulated using two-dimensional SBEs. (a) With an interband decoherence time of $T_2 = 6.6~\mathrm{fs}$, a slight negative shift in the delay corresponding to the maximum signal is observed. (b) With a longer decoherence time of $T_2 = 26~\mathrm{fs}$, the maximum shifts slightly to positive delay (unlike the exp. result).}\label{2DSBE_H4_delay}
\end{figure}

The 2D SBE simulations yielded qualitatively similar results (see Fig.~\ref{2DSBE_H4_delay}(a)) to the 1D case, i.e. there was a negative time shift for two-color delay leading to the maximum emission of the 4th harmonic. This result justifies applying the simpler model that is easier to analyze and is much more numerically tractable. We also note that we found the 2D model much more sensitive to $T_2$ dephasing time, whereby longer dephasing times caused the maximizing two-color delay to shift to positive values (see Fig.~\ref{2DSBE_H4_delay}(b)). This did not occur in the 1D model, or under any conditions in our experiments, and we believe might be a result of the $T_2$ phenomenological dephasing being a bad approximation in our conditions (where the total simulation duration far exceeds several tens of femtoseconds and either k-dependent dephasing\cite{korolev2024unveiling} or electron-phonon dynamics might be required \cite{lively2024revealing}).

\begin{figure}[htb!]
\centering
\includegraphics[width=0.6\textwidth]{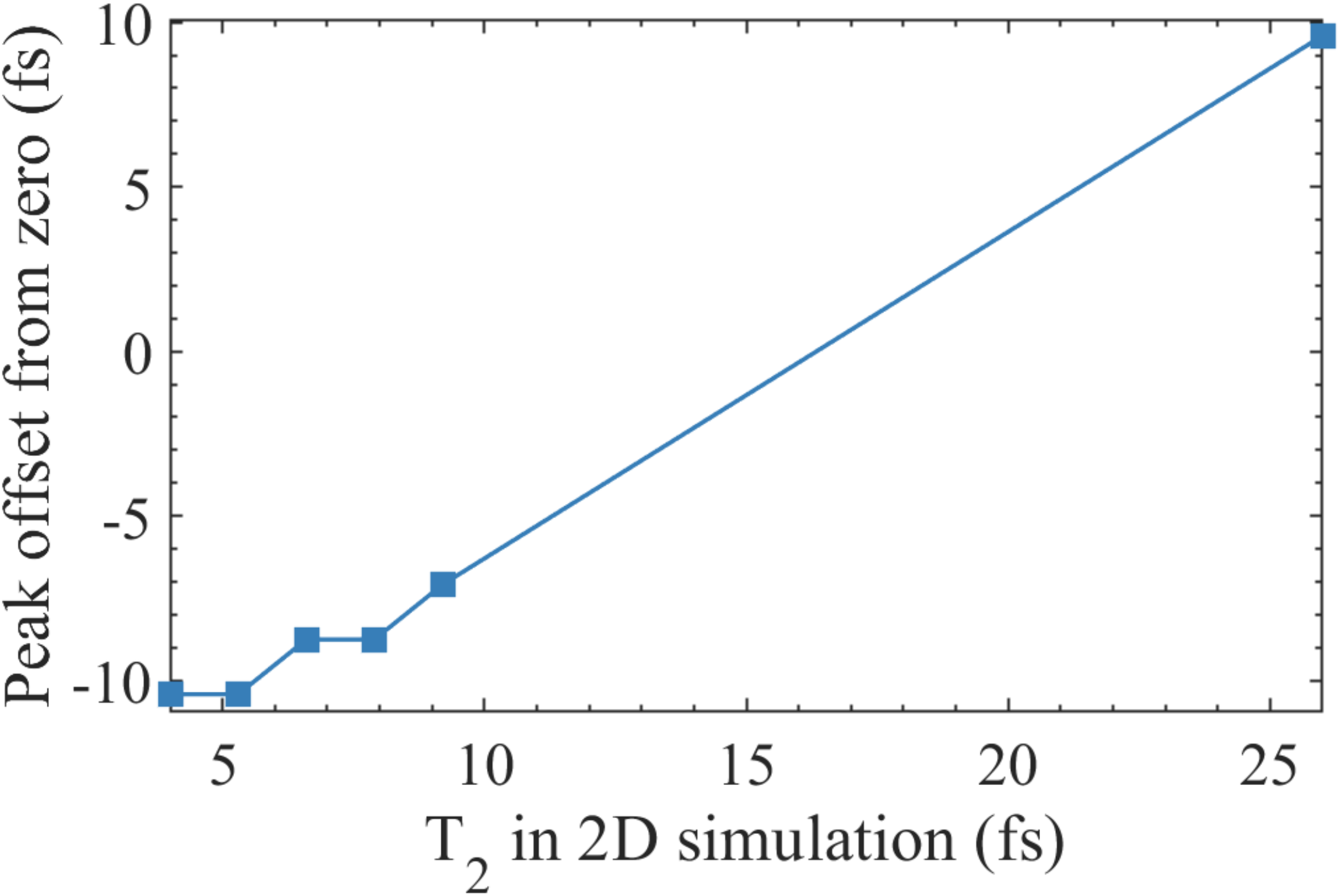} 
\caption{Negative shift of the H4 maximum in the $\omega$--$2\omega$ delay scanning predicted by the 2D graphene SBE model as a function of the decoherence time $T_2$.
}\label{2DSBE_T2_scan}
\end{figure}

We further calculate the delay of the peak of the fourth harmonic as a function of the decoherence time $T_2$, as shown in Fig. \ref{2DSBE_T2_scan}. For long decoherence times ($T_2 \gtrsim 16~\mathrm{fs}$), the peak position is slightly shifted to positive delays. As $T_2$ is reduced, the delay crosses zero around $T_2 \approx 16~\mathrm{fs}$ and becomes negative. Once $T_2$ is shorter than approximately one optical cycle of the fundamental field (about 6.6~fs at 2~\textmu m), the predicted delay becomes essentially insensitive to further changes in $T_2$.
\newpage

\subsection*{DFT-calculated band structures of graphene and HOPG}

\begin{figure}[htb!]
\centering
\includegraphics[width=0.9\textwidth]{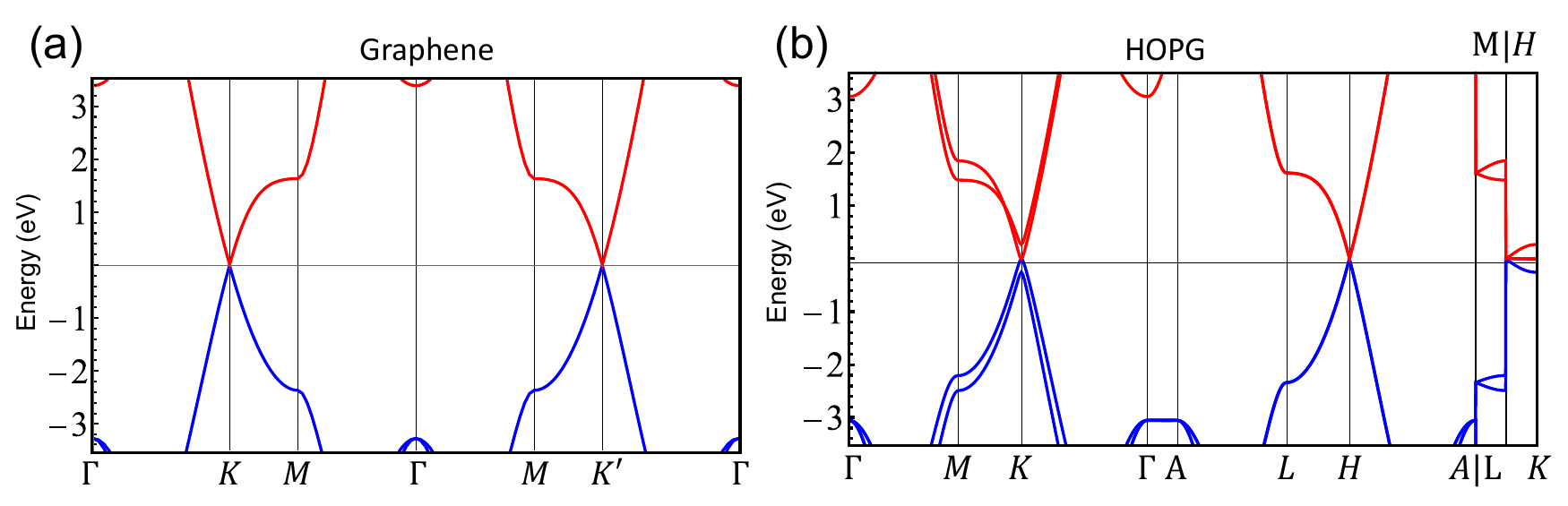} 
\caption{DFT-calculated electronic band structures along high-symmetry directions. (a) Monolayer graphene; (b) Highly oriented pyrolytic graphite (HOPG).  }\label{SM_2D_band}
\end{figure}

Here, we present the electronic band structures of graphene and HOPG along high-symmetry paths, as shown in Fig.~\ref{SM_2D_band}, calculated using first-principles DFT (following the procedures in ref.~\cite{neufeld2023band}). Notably, HOPG exhibits a Dirac cone-like dispersion near the $H$ point, analogous to the well-known Dirac cone at the $K$ or $K'$ point in monolayer graphene. However, in AB-stacked HOPG, interlayer coupling breaks the original symmetry of the monolayer, leading to a distortion and partial lifting of the Dirac cone structure.

\subsection*{Blue-shift of harmonic spectrum}
\begin{figure}[htb!]
\centering
\includegraphics[width=0.9\textwidth]{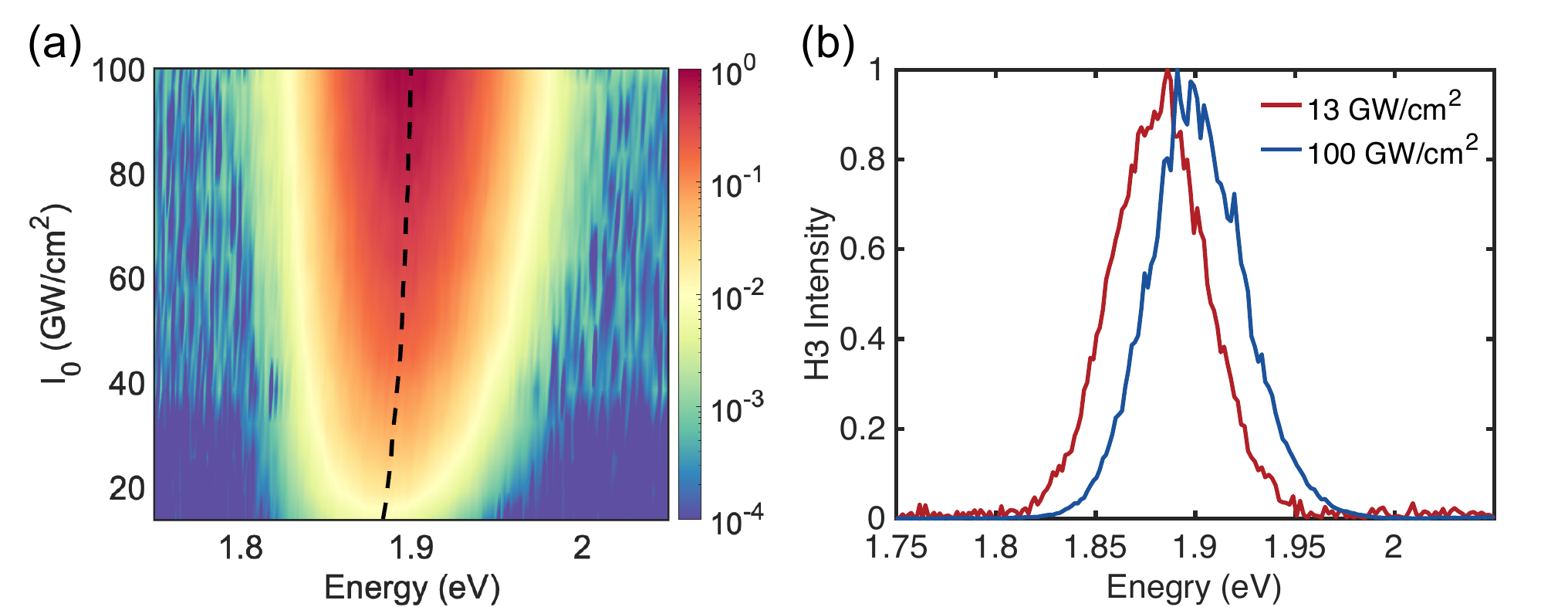} 
\caption{Blue-shift of the harmonic spectrum.  (a) Harmonic spectra as a function of the fundamental pump intensity. The black dashed line indicates the spectral center of H3. (b) Comparison of the H3 spectra obtained at two different pump intensities.}\label{H3_blueshift}
\end{figure}

We also observe a blue shift of the third-harmonic spectrum with increasing pump intensity, as shown in Fig.~\ref{H3_blueshift}. A direct comparison of the H3 spectra at pump intensities of $13\mathrm{GW/cm^2}$ and $100~\mathrm{GW/cm^2}$ shows a spectral shift of about 0.02 eV. This trend is analogous to the well-known blue shift in atomic HHG. In that case, the harmonic emission is intrinsically negatively chirped: the blue components are generated predominantly on the rising edge of the driving pulse, whereas the red components arise mainly from the falling edge. At high ionization or saturation regime, emission from the falling edge is strongly suppressed due to depletion, which reduces the red contribution. As a result, the spectrum becomes dominated by the blue-shifted components, leading to an overall net blue shift of the emitted harmonics~\cite{Gaarde2008,salieres1999study}.

\subsection*{Comparison with hexagonal WS$_2$}
\begin{figure}[htb!]
\centering
\includegraphics[width=0.7\textwidth]{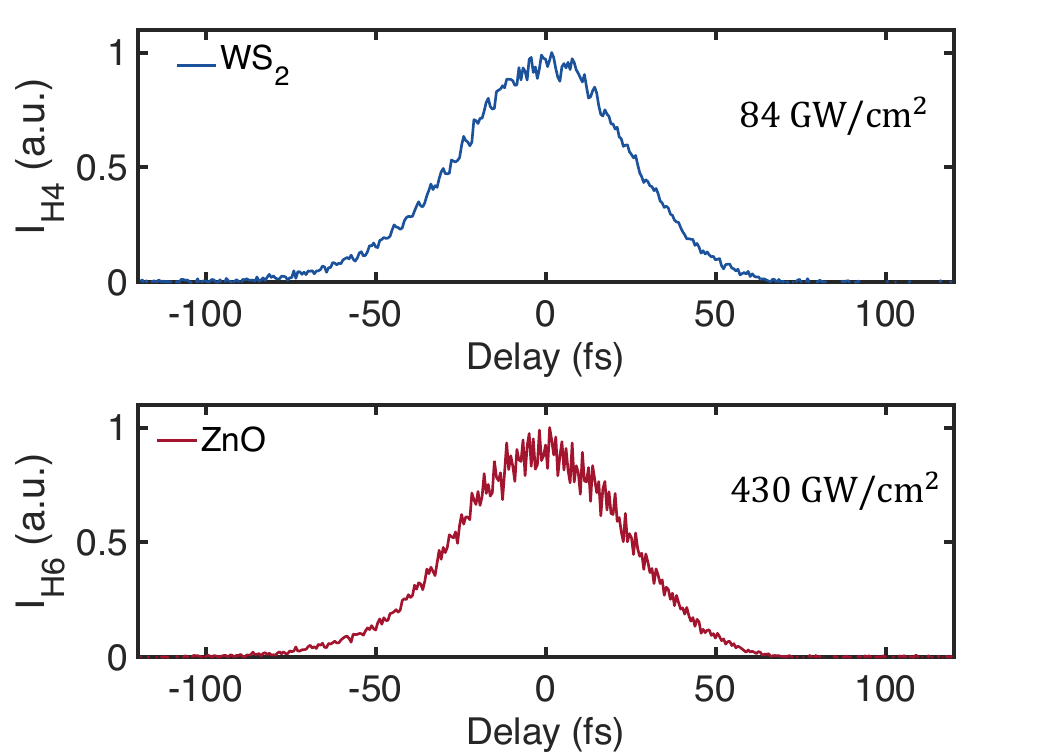} 
\caption{Above-bandgap harmonics from WS$_{2}$ (H4) and ZnO (H6) as a function of the two-color pump delay. The fundamental pump intensities are $84 ~\mathrm{GW/cm}^2$ for WS$_{2}$ and $430 ~\mathrm{GW/cm}^2$ for ZnO. In both cases, the harmonic yield peaks around zero delay, and no pronounced negative shift of the maximum is observed.}\label{H4_WS2}
\end{figure}

It is also natural to consider hexagonal wide-bandgap materials such as transition metal dichalcogenides (TMDs) as reference systems. To address this point, we performed additional measurements on WS$_2$ and compared the results with ZnO. Figure~\ref{H4_WS2} shows above-bandgap harmonics from WS$_2$ (H4) and ZnO (H6) as functions of the two-color pump delay, with fundamental intensities of 84~GW\,cm$^{-2}$ and 430~GW\,cm$^{-2}$, respectively. In both materials, the harmonic yield peaks around zero delay and exhibits no pronounced negative shift of the maximum, in stark contrast to the behavior observed in HOPG. These observations confirm that wide-bandgap hexagonal semiconductors such as WS$_2$ behave consistently with ZnO and remain far from strong depletion under our experimental conditions, thereby supporting our interpretation that the early-time harmonic peak in HOPG originates from saturation-driven dynamics in the gapless Dirac cone.

\subsection*{Influence of penetration depth in $z$-direction}
Since our measurements are performed in reflection geometry, only a near-surface region of the sample contributes to the detected harmonics. Using the optical constants of graphite reported by Querry \cite{querry1985optical}, the absorption coefficient at a wavelength of 2 µm is $\alpha \approx 12.3,\mathrm{m^{-1}}$, corresponding to a $1/e$ intensity penetration depth of $\delta \approx 1/\alpha \approx 80$ nm. This effective thickness is much smaller than both the pump wavelength (2 µm) and the wavelengths of the detected harmonics. Over such a short distance, the accumulated phase of both the pump and the harmonics amounts to only a small fraction of an optical cycle, while the driving field amplitude is progressively attenuated. As a result, propagation and longitudinal phase-matching effects in the bulk are strongly suppressed in our configuration, and the measured harmonics can be regarded as originating from a thin near-surface layer.

\begin{figure}[htb!]
\centering
\includegraphics[width=0.5\textwidth]{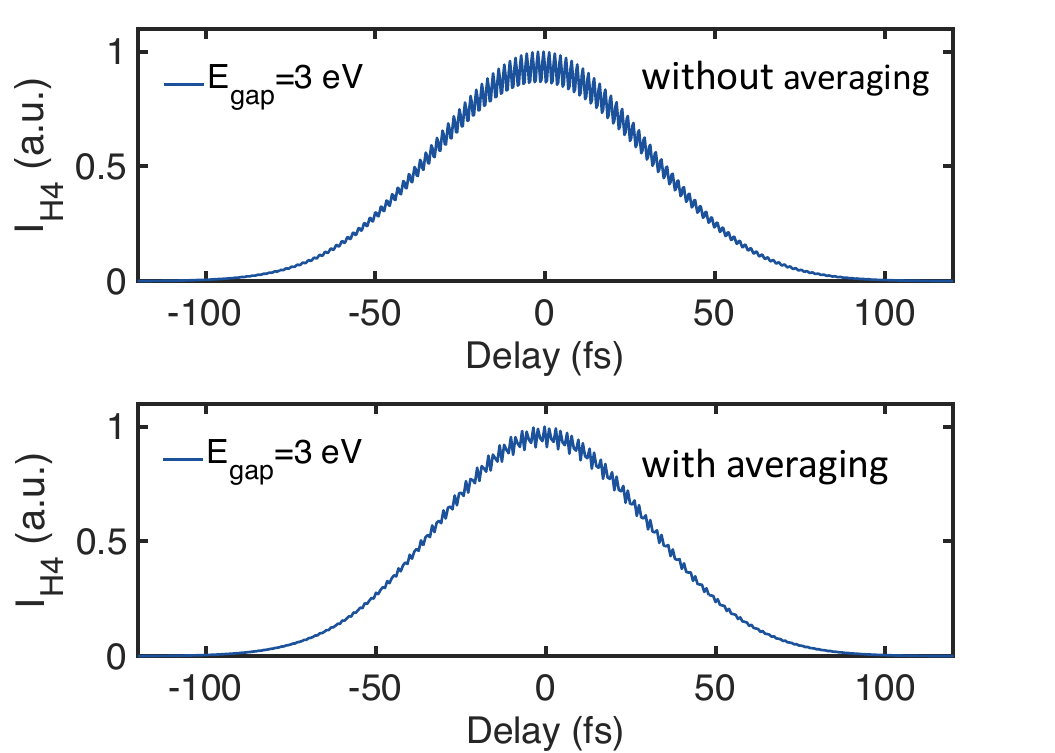} 
\caption{Comparison of H4 intensity in the $\omega-2\omega$ delay scanning in a gap system with and without averaging in penetration depth.}\label{Gap_Z}
\end{figure}

\begin{figure}[htb!]
\centering
\includegraphics[width=0.5\textwidth]{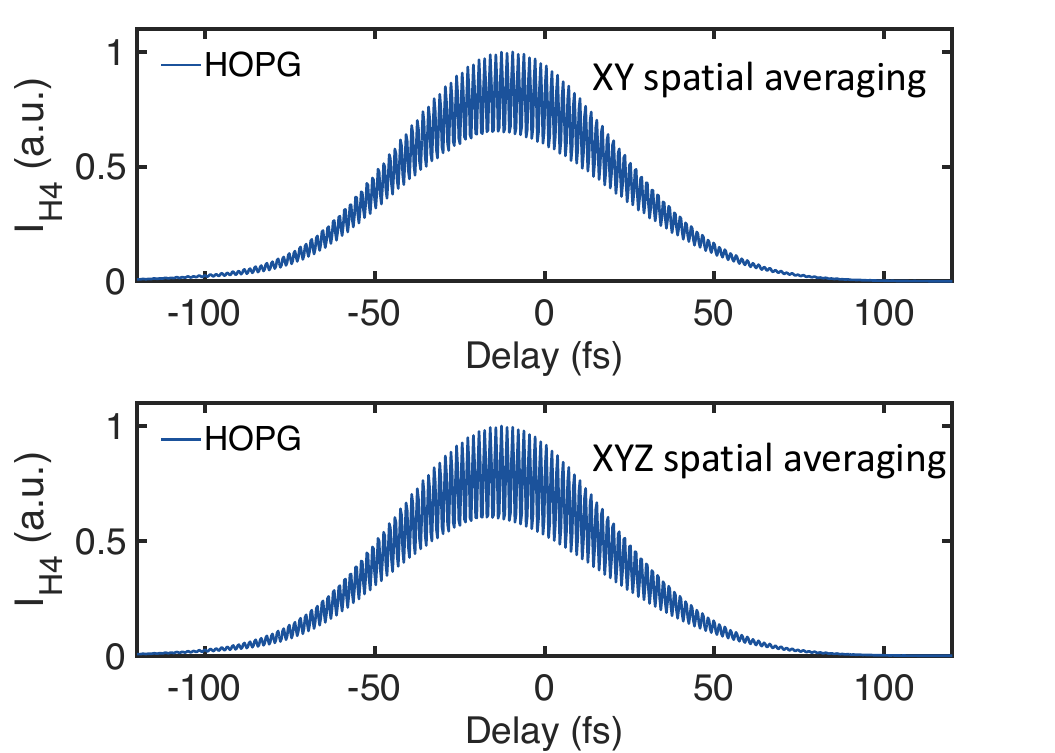} 
\caption{Comparison of H4 intensity in the $\omega-2\omega$ delay scanning in a gapless Dirac system with and without averaging in penetration depth.}\label{Gapless_Z}
\end{figure}

To test whether interference between contributions from different depths could nevertheless give rise to an apparent temporal shift, we simulated harmonic emission from layers distributed over a penetration depth of 80 nm and coherently averaged the resulting fields. Specifically, the depth-averaged harmonic intensity can be written as
\begin{equation}
I_{\mathrm{HH}} = \left| \int_{0}^{\infty} E_{\mathrm{HH}}\!\big(I(z')\big)\, e^{i k_{\mathrm{HH}} z'} \, \mathrm{d}z' \right|^{2},
\label{z_averaging}
\end{equation}
where $I(z') = I_{0}\exp(-\alpha z')$ is the depth-dependent pump intensity and $E_{\mathrm{HH}}$ is the corresponding harmonic electric field. As a control, we first introduced an artificial bandgap of 3 eV, for which no depletion occurs under 2 µm driving. In this gapped case, depth averaging reduces the modulation contrast but does not generate a negative delay (see Fig.~\ref{Gap_Z}), demonstrating that interference alone cannot account for the experimentally observed shift. We then repeated the calculation for the gapless Dirac cone and included Gaussian spatial averaging in the transverse ($x$–$y$) plane. Again, depth averaging has only a minor influence on the extracted delay, which remains robustly negative in the two-color scans (Fig.~\ref{Gapless_Z}), confirming that the observed temporal shift originates from genuine strong-field carrier dynamics rather than from propagation-induced interference effects.

\subsection*{Theoretical model for high harmonic generation from HOPG}

The modeling of the 3D SBE for HOPG followed the 2D scheme described above, but where $z$-axis dispersion was added into simulations (with additional 20 $k$-point sampling along $k_z$, which converged the HHG emission). The SBE were propagated in the basis of a 4x4 tight-binding (TB) Hamiltonian considering only nearest-neighbor terms for the interlayer hopping, and only up to 2nd nearest-neighbor terms within the individual planes for simplicity. However, we found that the role of additional in-plane nearest-neighbor terms did not affect the qualitative results. This 4x4 Hamiltonian was analytically diagonalized for the laser-free system, from which the dipole and momentum matrix elements were calculated to analytical form and numerically softened near K/K' (just as in the 2D case). We set the interlayer hopping parameter by fitting the gap opening of the TB model between bands \#1 and \#4 at K to values obtained from DFT simulations described above, leading to an interlayer hopping value of 0.0127\,eV. Simulations otherwise followed the same SBE scheme discussed above, where we propagated the 4x4 density matrix assuming full initial occupation in the bottom two valence bands, and with a similar $T_2$ phenomenological dephasing,where the laser was polarized within the graphite planes. By setting the interlayer hopping to zero, the 3D simulation effectively reproduces the 2D case for monolayer graphene. Propagation was achieved with 4th-order Runge-Kutte propagator. Additional technical details of this approach will soon be published in a separate work describing the model~\cite{Uzner2026}.

\newpage


\newpage

\newpage

\newpage

\newpage

\newpage

\newpage

\newpage

\newpage

\newpage

\newpage

\newpage

\newpage

\newpage

\newpage

\newpage


